\documentclass[notitlepage,apj,numberedappendix]{emulateapj}

\usepackage{verbatim}
\usepackage{amsmath,amsbsy}
\usepackage{hyperref}
\usepackage{breakurl}
\usepackage{float}
\usepackage{color}

\newcommand{\simgt}{\,\hbox{\lower0.6ex\hbox{$\sim$}\llap{\raise0.6ex\hbox{$>$}}}\,}
\newcommand{\simlt}{\,\hbox{\lower0.6ex\hbox{$\sim$}\llap{\raise0.6ex\hbox{$<$}}}\,}

\newcommand{\code}[1]{\texttt{#1}}

\newcommand*\Bell{\ensuremath{\boldsymbol\ell}}

\begin{document}

\title{What Do the {\it Hitomi} Observations Tell Us About the Turbulent Velocities in the Perseus Cluster? Probing the Velocity Field with Mock Observations}
  
\author{J. A. ZuHone\altaffilmark{1}, E. D. Miller\altaffilmark{2}, E. Bulbul\altaffilmark{2}, I. Zhuravleva\altaffilmark{3,4}}

\altaffiltext{1}{Harvard-Smithsonian Center for Astrophysics, 60 Garden St., Cambridge, MA 02138, USA}
\altaffiltext{2}{Kavli Institute for Astrophysics and Space Research, Massachusetts
Institute of Technology, 77 Massachusetts Avenue, Cambridge, MA 02139, USA}
\altaffiltext{3}{Kavli Institute for Particle Astrophysics and Cosmology,
  Stanford University, 452 Lomita Mall, Stanford, California 94305-4085, USA}
\altaffiltext{4}{Department of Physics, Stanford University, 382 Via Pueblo Mall,
  Stanford, California 94305-4060, USA}


\begin{abstract}
{\it Hitomi} made the first direct measurements of galaxy cluster gas motions in the Perseus cluster, which implied that its core is fairly ``quiescent'', with velocities less than $\sim$200~km~s$^{-1}$, despite the presence of an active galactic nucleus and sloshing cold fronts. Building on previous work, we use synthetic {\it Hitomi}/SXS observations of the hot plasma of a simulated cluster with sloshing gas motions and varying viscosity to analyze its velocity structure in a similar fashion. We find that sloshing motions can produce line shifts and widths similar to those measured by {\it Hitomi}. We find these measurements are unaffected by the value of the gas viscosity, since its effects are only manifested clearly on angular scales smaller than the SXS $\sim$1' PSF. The PSF biases the line shift of regions near the core as much as $\sim 40-50$~km~s$^{-1}$, so it is crucial to model this effect carefully. We also infer that if sloshing motions dominate the observed velocity gradient, Perseus must be observed from a line of sight which is somewhat inclined from the plane of these motions, but one that still allows the spiral pattern to be visible. Finally, we find that assuming isotropy of motions can underestimate the total velocity and kinetic energy of the core in our simulation by as much as $\sim$60\%. However, the total kinetic energy in our simulated cluster core is still less than 10\% of the thermal energy in the core, in agreement with the {\it Hitomi} observations.
\end{abstract}

\section{Introduction}\label{sec:intro}

The dominant baryonic component of galaxy clusters, the intracluster medium (ICM), emits prodigiously in X-rays. Coupled with significant theoretical progress in understanding the underlying emission mechanisms, the past and current generation of X-ray telescopes, especially {\it Chandra}, {\it XMM-Newton}, and {\it Suzaku}, have revealed a wealth of knowledge about the properties of the ICM, including its density, temperature, and chemical composition \citep{rei09,eck13,mer15,mcd16,bar17,eze17}

Since clusters of galaxies are dynamic objects, forming as the result of the bottom-up process of cosmological structure formation, the kinematical properties of the ICM are also important. Theoretical studies of galaxy clusters have shown that determining the properties of the ICM velocity field is important for a number of reasons. Kinetic energy in the form of bulk motions and turbulence provides a form of pressure support against gravity supplemental to thermal pressure, biasing mass estimates based on the assumption of hydrostatic equilibrium, as predicted by simulations \citep{evr96,ras06,nag07,pif08,tak10,sut13,nel14}. Dissipation of turbulent kinetic energy into heat, in addition to turbulent transport and mixing of hot gas, may partially offset gas cooling in cluster cool cores \citep{fuj04,den05,zuh10,ban14,zhu14}. The velocity structure on small length scales places constraints on the microphysics of the ICM, in particular its viscosity \citep{fab03,rod13,zuh15}. Finally, ICM turbulence is likely a key ingredient for the origin of non-thermal phenomena such as radio halos and radio mini-halos \citep{ohn02,bru07,don13,zuh13,fuj15}.

\begin{figure*}
\begin{center}
\includegraphics[width=0.9\textwidth]{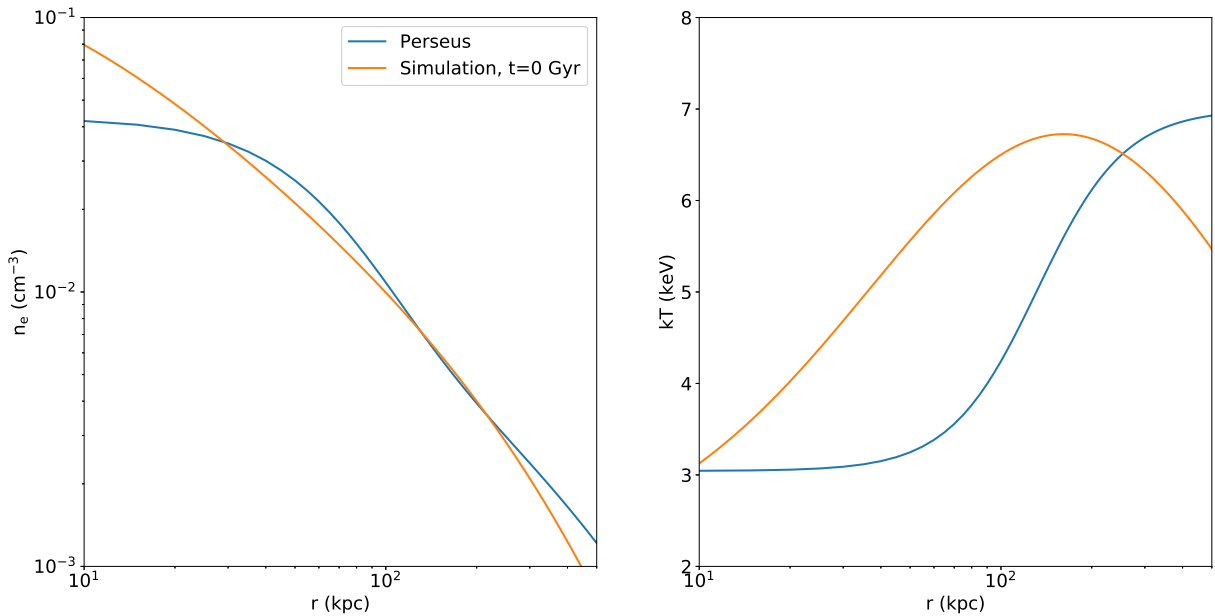}
\caption{Density and temperature profiles of our cluster model compared to those from the Perseus cluster, using the analytical fitting formulas from \citet{chu03}.\label{fig:cluster_profiles}}
\end{center}
\end{figure*}

However, up until recently the kinematical properties of the cluster plasma were largely elusive, due to the fact that no X-ray instrument had the spectral resolution required to resolve shifting and broadening of spectral lines due to the Doppler effect. Nearly all previous indications of motions in the ICM, though largely indisputable, had been indirect. The RGS grating on {\it XMM-Newton} can provide weak upper limits on Doppler broadening of spectral lines in cool-core clusters \citep[][and references therein]{san11,bul12,san13,pin15}. Upper limits on line shifts in the ICM were also determined with the X-ray Imaging Spectrometer (XIS) on {\it Suzaku} \citep[][]{ota07,sug09,tam14}. In one cluster, A2256, the bulk motion was fast enough ($\sim$1500~km~s$^{-1}$) to produce a line shift measurable by XIS \citep[][]{tam11}. Indirect estimates of the ICM turbulent velocity can be obtained from measurements of resonant scattering \citep[e.g.,][]{chu04,wer09,dep12,zhu13}, pressure fluctuations \citep{sch04,kha16}, or surface brightness fluctuations \citep{chu12,zhu15}. Also, the existence of features such as shock fronts and cold fronts are clear indications of gas motions driven by cluster mergers \citep[][]{MV07} and active galactic nucleus (AGN) activity \citep[][]{ran15}.

The general possibility of measuring gas motions in galaxy clusters directly was first achieved recently by the {\it Hitomi} X-ray Observatory \citep[][]{tak14}. {\it Hitomi} possessed a Soft X-ray Spectrometer (SXS) microcalorimeter with an energy resolution of $\Delta{E} \sim 4.5$~eV within the energy range $E \sim 0.3-12.0$~keV, covering a 3'$\times$3' field. At the energy of the Fe-K$_{\alpha}$ line, $E \approx 6.7$~keV, the SXS spectral resolution enabled the measurement of velocities at resolutions of tens of km~s$^{-1}$. Sadly, in late March of 2016 {\it Hitomi} lost contact with the ground, and it was unable to be recovered.

{\it Hitomi} observed the core of the Perseus cluster (Abell 426) in early 2016 with the SXS. The analysis of two observations with a total of 230~ks of exposure time were reported in \citet[][hereafter H16]{hit16} The analysis of two additional observations, for a combined total of 320~ks of exposure time, were reported in \citet[][hereafter H17]{hit17a}. The Perseus cluster is an ideal candidate to study gas motions in clusters. First, it is nearby ($z$ = 0.0179), large, and bright. The central galaxy, NGC~1275, possesses a powerful AGN that is blowing bubbles into the Perseus ICM, driving shocks, turbulence, and sound waves \citep[][]{boe93,chu00,fab00,fab02,fab03,fab06,zhu16}. Additionally, the spiral-shaped cold fronts beginning in the core region and extending out to larger radii indicate the presence of sloshing gas motions (see the left panel of Figure 3 of H16), presumably initiated by a previous encounter with a subcluster \citep{chu03,fab11,sim12,wal17}.

Both H16 and H17 reported the measurement of line shifts and broadening in Perseus, at a significance which clearly indicates the presence of gas motions. However, H16 reported the gas motions in the core were somewhat modest, with a line-of-sight velocity dispersion of 164$\pm$10~km~s$^{-1}$, and a gradient in the line-of-sight velocity of 150$\pm$70~km~s$^{-1}$. The implied pressure support from the velocity dispersion is $\sim$4\%, and in combination with the contribution from bulk motions it is still less than 10\%. H17 reported results that were consistent with this, showing that near the AGN and ``ghost'' bubble to the northwest the velocity dispersion is $\sim$200~km~s$^{-1}$, whereas elsewhere the velocity dispersion is lower at $\sim$100~km~s$^{-1}$. They also reported a bulk velocity gradient across the core region of $\sim$100~km~s$^{-1}$. In a separate paper, \citet{hit17b} presented evidence for resonant scattering in the core of Perseus based on the {\it Hitomi} measurements, obtaining estimates on turbulent velocities consistent with those measured from line-of-sight Doppler shifts. Given that {\it Chandra} and {\it XMM-Newton} observations of Perseus indicate the presence of ``cluster weather'' due to AGN activity and gas sloshing, the apparently ``quiescent'' nature of the cluster core indicated by the {\it Hitomi} observations comes as somewhat of a surprise. The authors of H16 noted that ``a low level of turbulent pressure measured for the core region of a cluster, which is continuously stirred by a central AGN and gas sloshing, is surprising and may imply that turbulence in the intracluster medium is difficult to generate and/or easy to damp'' (page 119, H16).

For the reasons listed above, the discovery of such a low level of gas motion has important implications. If gas motions are difficult to generate or easy to damp, it may imply that a) the sources of cluster weather are not as strong as previously thought or b) that the viscosity of the cluster gas may be significant, potentially providing constraints on the plasma physics of the ICM and impacting turbulent reacceleration models for radio halos and radio mini-halos. For this reason, it is important to determine what implications the {\it Hitomi} observations of Perseus may have for these questions. 

\begin{figure*}
\begin{center}
\includegraphics[width=0.96\textwidth]{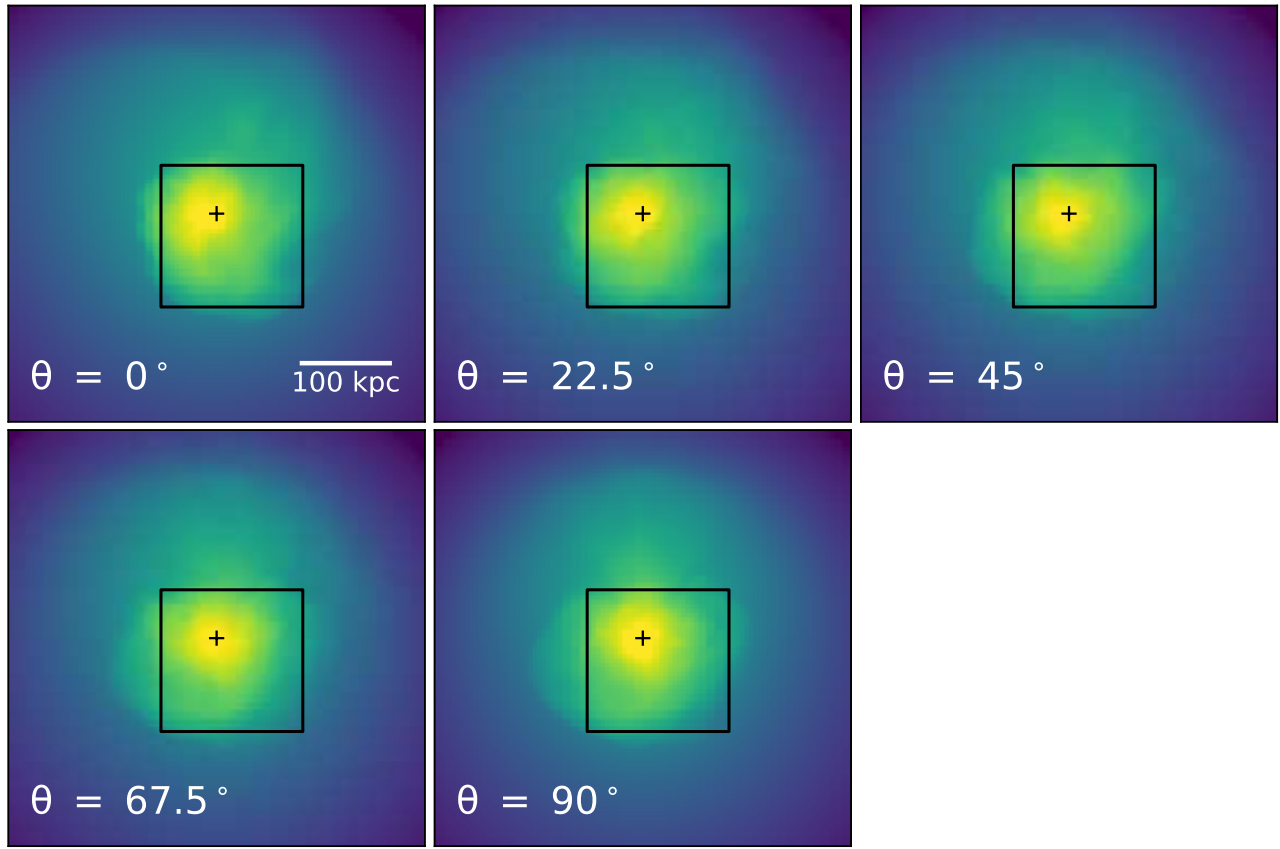}
\caption{Maps of the projected X-ray surface brightness in the 0.6-9~keV band of the inviscid simulation, projected along 5 different lines of sight, each given in terms of the angle between the $z$ and the $x$-axis of the simulation. The cross indicates the position of the gravitational potential minimum of the cluster, and the square indicates the location of the simulated SXS pointing.\label{fig:novisc_S_X}}
\end{center}
\end{figure*}

\begin{figure*}
\begin{center}
\includegraphics[width=0.96\textwidth]{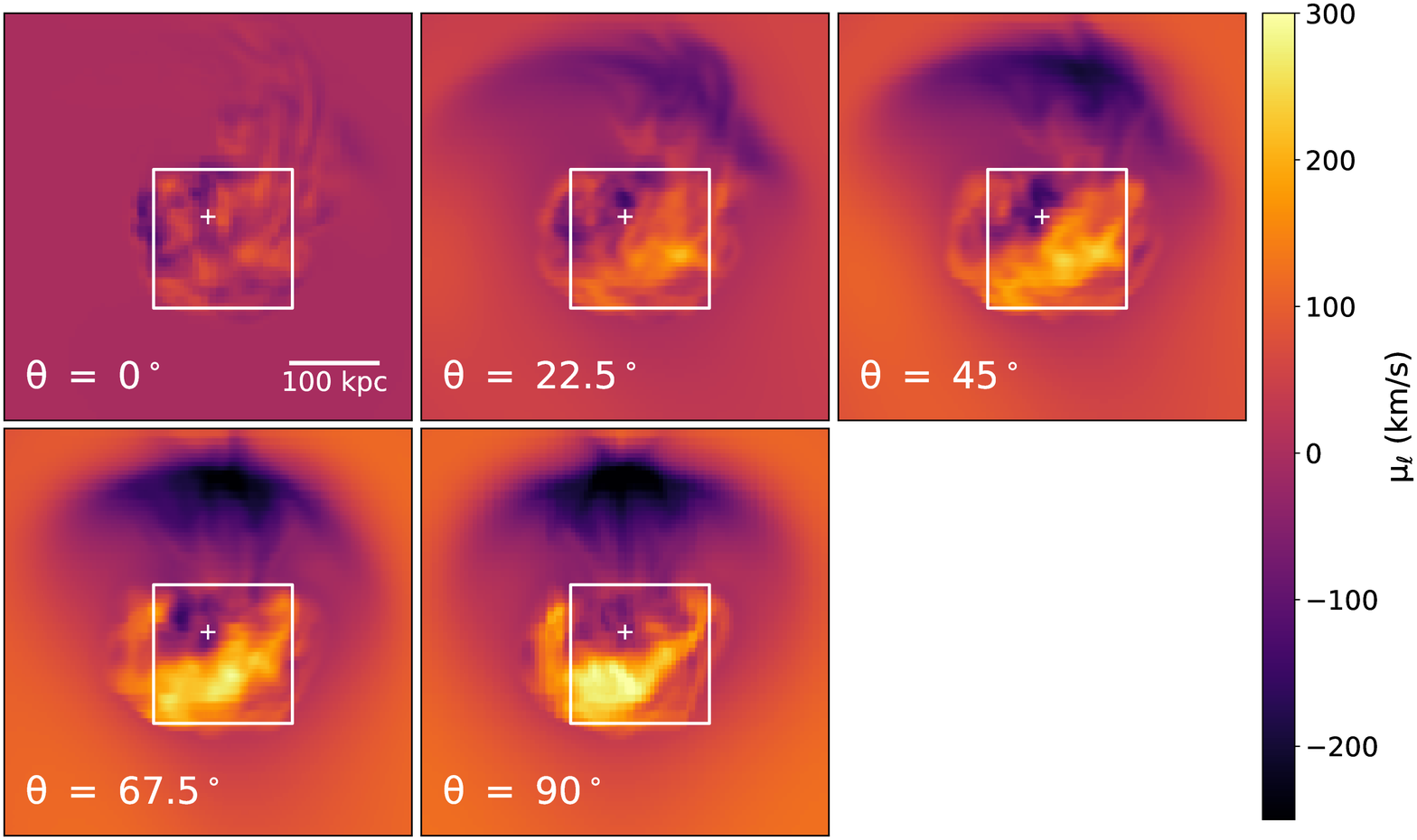}
\caption{Maps of the emission-weighted velocity line shift in km~s$^{-1}$ of the inviscid simulation, projected along 5 different lines of sight, each given in terms of the angle between the $z$ and the $x$-axis of the simulation. The cross indicates the position of the gravitational potential minimum of the cluster, and the square indicates the location of the simulated SXS pointing.\label{fig:novisc_mu}}
\end{center}
\end{figure*}

\begin{figure*}
\begin{center}
\includegraphics[width=0.96\textwidth]{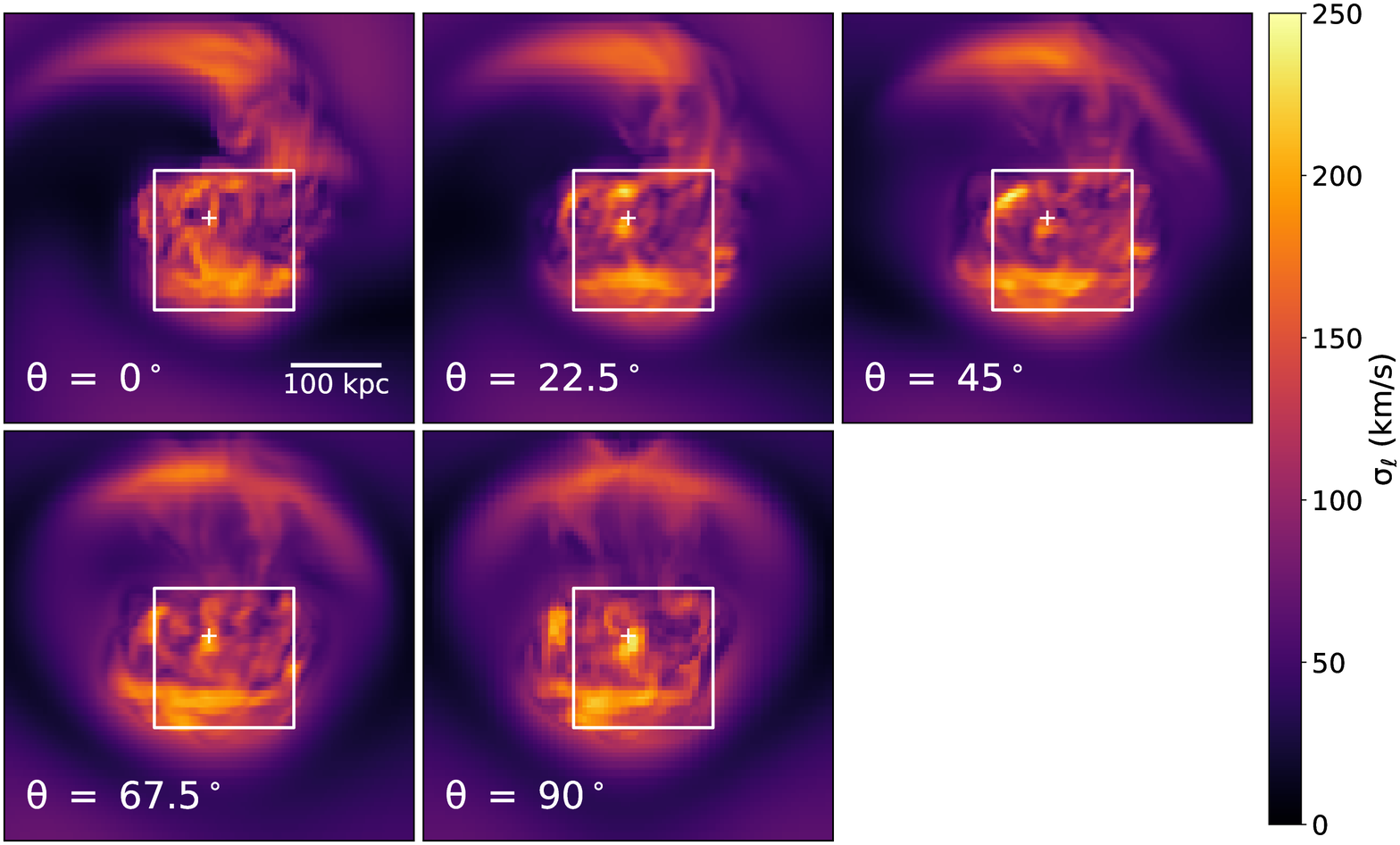}
\caption{Maps of the emission-weighted velocity line width in km~s$^{-1}$ of the inviscid simulation, projected along 5 different lines of sight, each given in terms of the angle between the $z$ and the $x$-axis of the simulation. The black cross indicates the position of the gravitational potential minimum of the cluster, and the square indicates the location of the simulated SXS pointing.\label{fig:novisc_sigma}}
\end{center}
\end{figure*}

\begin{figure*}
\begin{center}
\includegraphics[width=0.96\textwidth]{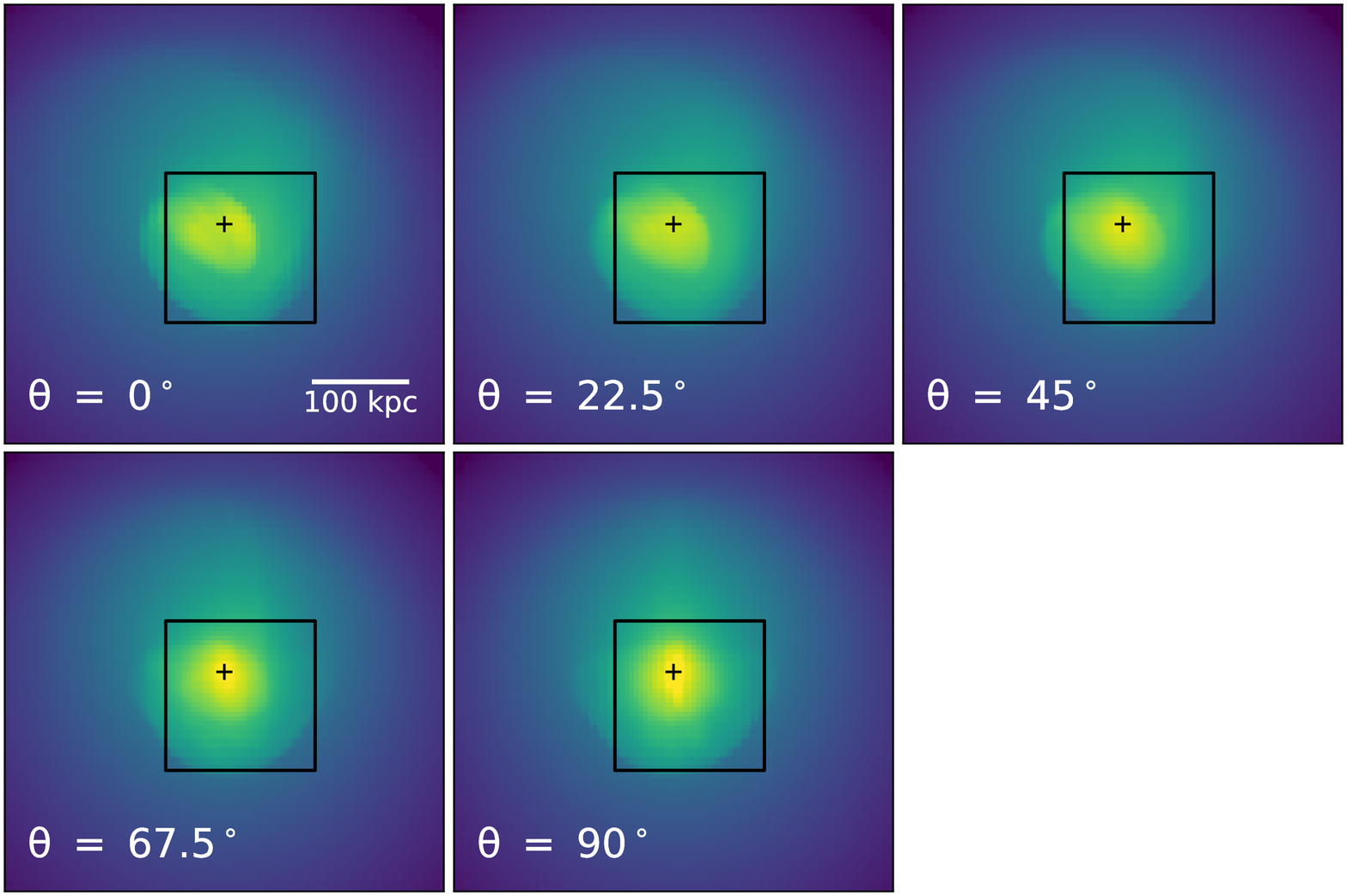}
\caption{Maps of the projected X-ray surface brightness in the 0.6-9~keV band of the viscous simulation, projected along 5 different lines of sight, each given in terms of the angle between the $z$ and the $x$-axis of the simulation. The cross indicates the position of the gravitational potential minimum of the cluster, and the square indicates the location of the simulated SXS pointing.\label{fig:visc_S_X}}
\end{center}
\end{figure*}

\begin{figure*}
\begin{center}
\includegraphics[width=0.96\textwidth]{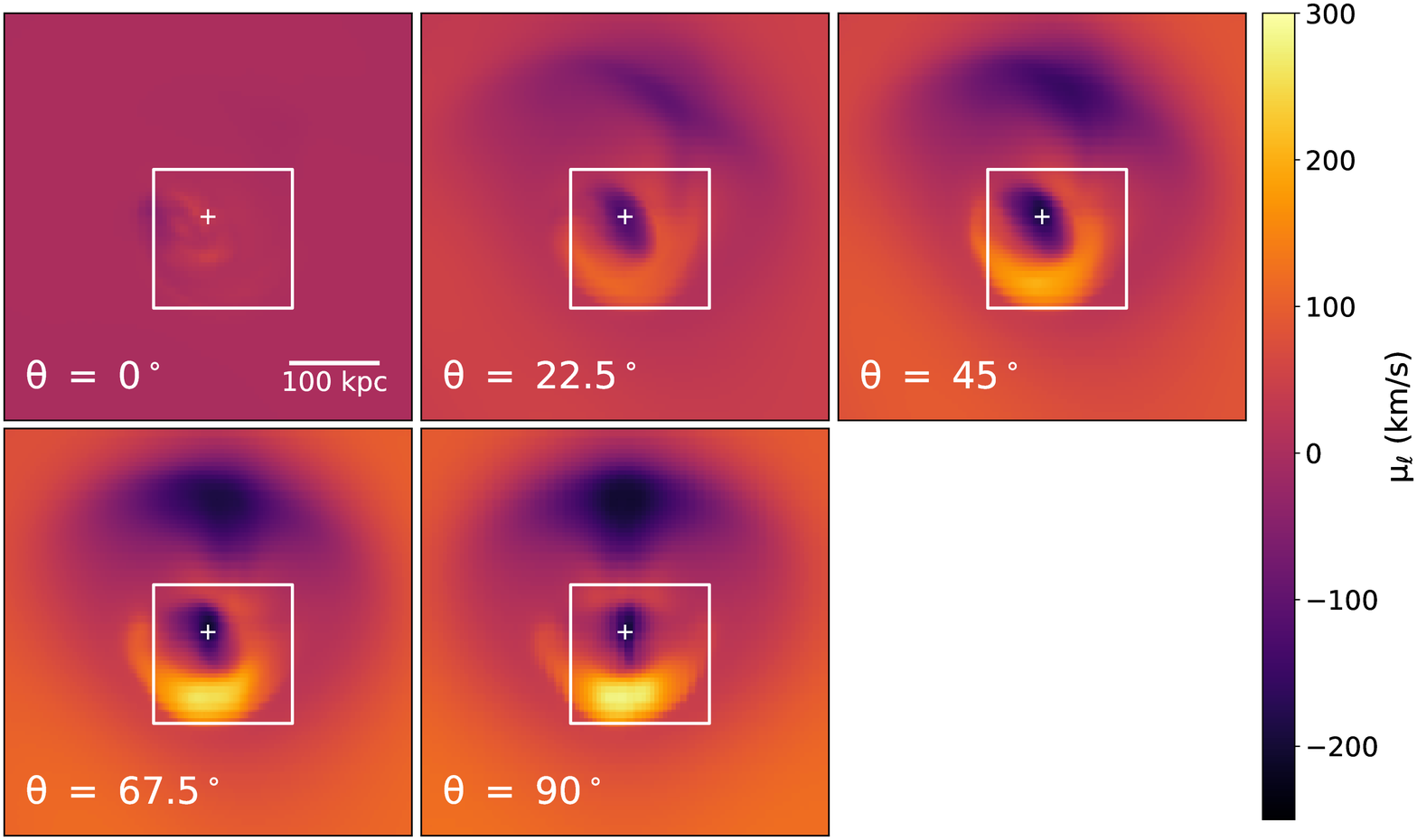}
\caption{Maps of the emission-weighted velocity line shift in km~s$^{-1}$ of the viscous simulation, projected along 5 different lines of sight, each given in terms of the angle between the $z$ and the $x$-axis of the simulation. The cross indicates the position of the gravitational potential minimum of the cluster, and the square indicates the location of the simulated SXS pointing.\label{fig:visc_mu}}
\end{center}
\end{figure*}

\begin{figure*}
\begin{center}
\includegraphics[width=0.96\textwidth]{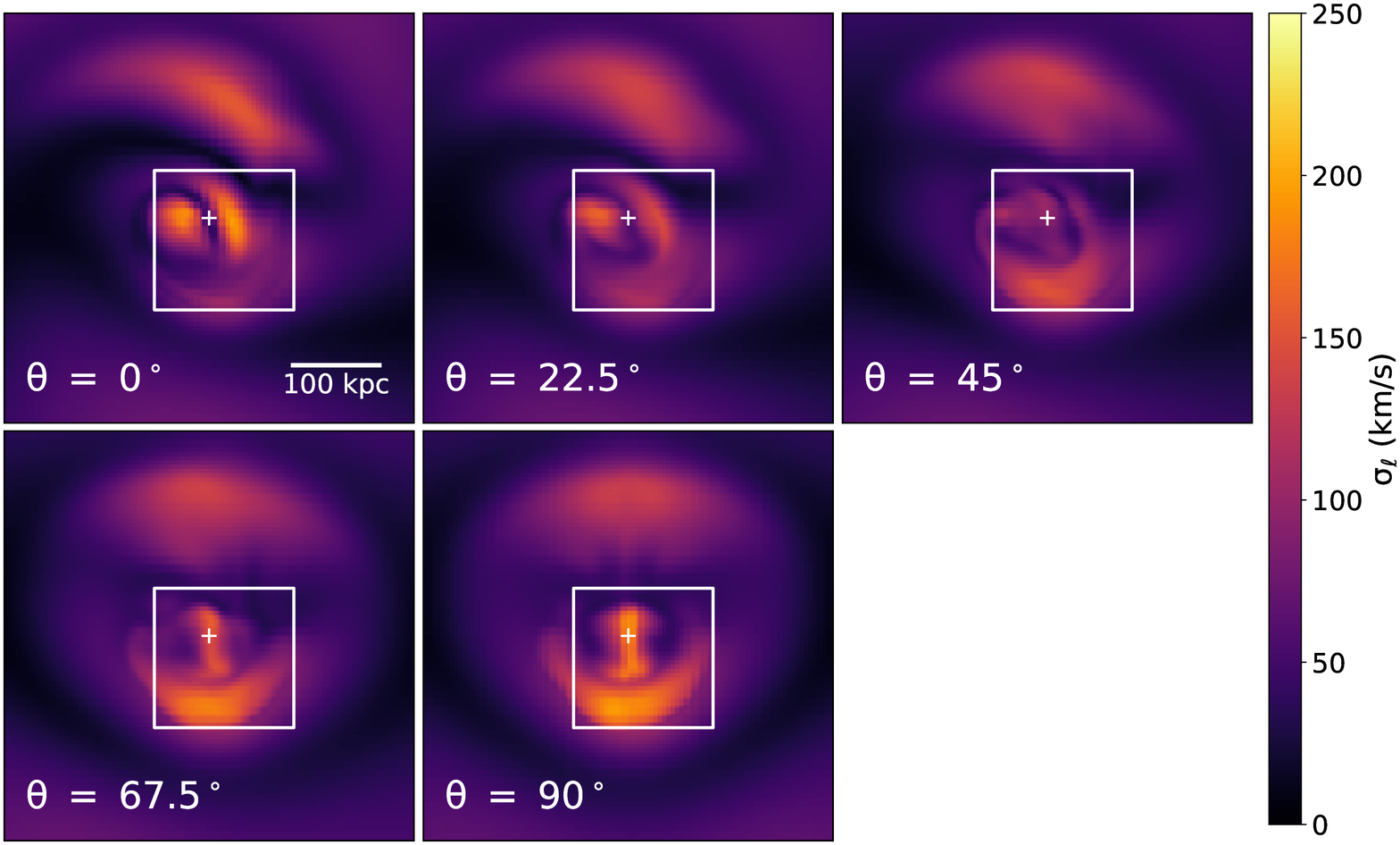}
\caption{Maps of the emission-weighted velocity line width in km~s$^{-1}$ of the viscous simulation, projected along 5 different lines of sight, each given in terms of the angle between the $z$ and the $x$-axis of the simulation. The cross indicates the position of the gravitational potential minimum of the cluster, and the square indicates the location of the simulated SXS pointing.\label{fig:visc_sigma}}
\end{center}
\end{figure*}

In this work, we use hydrodynamical simulations of gas sloshing in a galaxy cluster core similar to that of Perseus to investigate these questions. In particular, we seek to determine whether or not it is true that conclusions about the damping properties of the plasma, or its viscosity, can be drawn from the {\it Hitomi} observations. To do this, we examine the effect of viscosity on the measurements of gas motions in such a cluster by simulating two extreme cases: a cluster plasma which is inviscid and one which is very viscous, more than expected from theoretical arguments. We examine projected line shifts and widths directly from the simulation and compare them to those estimated from mock {\it Hitomi}/SXS observations, taking into account the spatial and spectral resolution of the instrument. An analysis of the velocity field of these simulations, including mock {\it Hitomi} observations, was previously presented in \citet[][hereafter Z16]{zuh16}, but in this work we seek to be informed by the {\it Hitomi} observations of Perseus by applying a very similar analysis and attempting to use them to place constraints on the light of sight along which Perseus is viewed from Earth. 

Of course, the core of Perseus hosts a powerful AGN, which will drive gas motions, and other sources of bulk and turbulent motion may exist within the core in addition to the sloshing motions evidenced by the spiral cold fronts. These possibilities have already been investigated by \citet{lau17}, \citet{bou17}, and \citet{hil17}. We will briefly comment on these results in Section \ref{sec:summary}. However, given the prominence of the spiral feature in Perseus, it is likely that its associated gas motions make a significant contribution to the observed line shifts and broadening seen in the {\it Hitomi} observations, and so it is worthwhile to examine it in isolation. 

The structure of this paper is as follows: in Section \ref{sec:methods} we briefly outline the setup of the galaxy cluster simulations and the procedure for creating mock X-ray observations. In Section \ref{sec:results} we present the results of our analysis, and in Section \ref{sec:summary} we summarize these results and present our conclusions. All calculations assume a flat $\Lambda$CDM cosmology with $h$ = 0.71, $\Omega_m$ = 0.27, and $\Omega_\Lambda$ = 0.73. 

\section{Methods}\label{sec:methods}

\subsection{N-body/Hydrodynamic Simulations}\label{sec:hydro_sims}

The two cluster merger simulations examined in this work were originally presented in \citet[][]{zuh10}, and model an off-center collision between a large, cool-core cluster and a smaller subcluster. This configuration produces sloshing cold fronts and gas motions in the large cluster's core. These simulations were performed with the parallel N-body/hydrodynamics adaptive mesh refinement (AMR) astrophysical simulation code \code{FLASH} \citep{dub09}. The full details of the setup of the simulations and algorithms employed can be found in \citet{zuh10} and Z16, but we provide a short summary here.

The simulations used \code{FLASH}'s standard hydrodynamics module employing the Piecewise-Parabolic Method of \citet{col84} for treatment of the cluster plasma, under the assumption of an ideal gas equation of state with $\gamma = 5/3$, and a mean molecular weight of $\mu = 0.592$, appropriate for an ionized H/He gas with a hydrogen mass fraction of $X = 0.75$. The dark matter component of the clusters is modeled by a collection of massive particles, using an $N$-body module which uses the particle-mesh method to map accelerations from the AMR grid to the particle positions. The gravitational potential of the self-gravitating gas and dark matter is computed using a multigrid solver \citep{ric08}. The physics of the two simulations are identical with the important exception that one is inviscid, and the other is viscous, with isotropic Spitzer viscosity \citep{spi62,sar88}:
\begin{eqnarray}\label{eqn:spitzer}
\mu &=& 0.960\,\frac{n_{\rm i}k_{\rm B}T}{\nu_{\rm ii}} \\
\nonumber &\approx& 2.2 \times 10^{-15}\frac{T^{5/2}}{\ln\Lambda_{\rm i}}~{\rm g~cm^{-1}~s^{-1}},
\end{eqnarray}
where $n_i$ is the ion number density, $\nu_{\rm ii}$ is the ion-ion collision frequency, the temperature $T$ is in Kelvin, and the ion Coulomb logarithm $\ln\Lambda_{\rm i} \approx 40$, appropriate for conditions in the ICM. Such a high viscosity for the ICM is unlikely, due to the anisotropic nature of the ion viscosity in a high-$\beta$ magnetized plasma \citep{bra65}, and also because microscale plasma instabilities may set an upper limit on the viscosity that is much lower than expected for a collisional plasma \citep{kun14}. However, this simulation still serves as a useful test case, since it allows us to examine the effects of the gas motions on the spectral lines in the limit that turbulence and instabilities are completely suppressed.

This work uses the simulations ``R5b500'' and ``R5b500v'' from \citet{zuh10}, the only difference between these two simulations being the addition of viscosity to the latter. Both simulations are set up with the initial condition of a large, $\sim10^{15}~M_\odot$ cool-core cluster, and a smaller, dark matter-only subcluster 5 times less massive, separated at a distance of 3~Mpc, with an impact parameter along the $y$-axis of the simulation of $b = 500$~kpc, on a bound orbit in the $x$-$y$ plane of the simulation domain. The initial velocities of the clusters are in the $x$-direction and are set using Equations 4 and 5 of \citet{zuh10}. Figure \ref{fig:cluster_profiles} shows the initial density and temperature profiles of our model cluster compared to analytical profiles fitted to the data of the Perseus cluster from \citet{chu03}. Though there are differences between our model and Perseus, the density and temperature profiles are very similar in terms of their shape and overall normalization, so our simulation is a good candidate for studying the dynamics of gas motions in a Perseus-like system. 

\subsubsection{Geometry of the Problem}\label{sec:geometry}

Due to the aforementioned symmetry of the simulation, the resulting gas motions are predominantly in the velocity components along the $x$ and $y$ axes, and the spiral pattern is seen most prominently in projections with lines of sight near the $z$-axis. For projected quantities and synthetic observations, we choose lines of sight which result in an appearance of the simulated cluster which closely resembles the position and orientation of the Perseus cold fronts. For all maps presented in this work, the ``up'' direction $\hat{\textbf{n}}$ corresponds to the -$y$-axis of the simulation, and the line-of-sight vector $\hat{\Bell}$ is determined by an angle $\theta$, which is the angle away from the $z$-axis of the simulation in the $x-z$ plane towards the $x$-axis. Therefore, an angle of $\theta = 0^\circ$ is aligned with the $z$-axis, and an angle of $\theta = 90^\circ$ is aligned with the $x$-axis.

As in Z16, we choose the epoch $t$ = 3.0~Gyr of both simulations, finding this to be a moment in time where the shape and orientation of the cold fronts is a good match to those in Perseus. However, our simulated cluster is not an exact match. In particular, the sizes of the cold fronts at this moment of the simulation are somewhat larger than those in Perseus, by a factor of roughly $\sim$2. For this reason, when making projections and synthetic observations we choose the redshift of the cluster to be $z = 0.043$ instead of the redshift $z = 0.0179$ of Perseus, in order that the size of the core region and the cold fronts as projected on the sky is comparable to the 3' width of the SXS field of view. For this reason, many of our comparisons to Perseus are qualitative. However, we stress that our general conclusions (detailed in Section \ref{sec:summary}) are likely to be very applicable to the interpretation of the {\it Hitomi} observations of Perseus. For the calculations in this work, we will work in the rest frame of the main cluster.

\subsection{Synthetic X-ray Observations}\label{sec:mock_obs_method}

The most important results from this work are derived from synthetic {\it Hitomi}/SXS observations of our simulated clusters. These observations are produced from our simulation outputs using the two software packages. We use the \code{pyXSIM}\footnote{\url{http://hea-www.cfa.harvard.edu/~jzuhone/pyxsim/}} \citep[][]{zuh14} package to create samples of X-ray photons from our hydrodynamic simulations, and we use the \code{SOXS}\footnote{\url{http://hea-www.cfa.harvard.edu/~jzuhone/soxs/}} package to convolve these X-ray photons with the {\it Hitomi}/SXS instrumental responses. 

\code{pyXSIM} takes a 3D hydrodynamic simulation and produces a distribution of synthetic X-ray photons from the simulation variables of density, temperature, and velocity assuming that the X-ray emission arises from a thermal plasma, using the PHOX algorithm originally described in \citet{bif12,bif13}. We assume the X-ray emission can be described by an \code{APEC} model and AtomDB version 3.0.8 \citep{smi01,fos12}. Since the simulation does not include metallicity, we assume a spatially constant metallicity of Z = 0.7~Z$_\odot$, appropriate for the Perseus core \citep[][]{mat11}. We assume \citet{asp09} abundances. Photons are generated from each cell in the simulation to produce a large initial sample of candidate events from which to produce mock observations. This sample is projected along several lines of sight $\hat{\Bell}$ to a 2D plane as described above. The energies of the photons are then Doppler shifted by the velocity $v_\ell = \hat{\Bell} \cdot \textbf{v}$ of their originating gas cells along the line of sight and cosmologically redshifted. Lastly, a number of photons are absorbed by foreground Galactic neutral hydrogen, assuming the Tuebingen-Boulder ISM absorption model \citep[\code{TBabs},][]{wil00}, assuming a Galactic column density of $N_H = 4 \times 10^{20}$~cm$^{-2}$. We do not explicitly include the effect of systematic errors due to the gain uncertainty of the SXS, which amounts to an error of approximately 50~km~s$^{-1}$ on line shift measurements. Therefore, the only source of error associated with our mock observation measurements is statistical. All error bars in figures are 1-$\sigma$, and errors on line shift, line width, and derived quantities have been computed via standard error propagation.

These photon samples then serve as inputs to the ``instrument simulator'' module of \code{SOXS}. We have implemented a simple model for the {\it Hitomi}/SXS instrument in \code{SOXS}, assuming a square field of view 3' on a side, with 0.5' pixels and a Gaussian spatial PSF of $\sim$1.2' half-power diameter. For estimating the effects of the PSF, we also use a separate configuration with no PSF which is otherwise identical. \code{SOXS} simulates the detection of the events, smears the position on the chip using the model for the PSF, and convolves the photon energies with an ARF and RMF that were created using the publicly available HEASOFT v6.20 FTOOLS, along with Hitomi CALDB v5 (release date 2016-12-23). We ignore the effects of instrumental and astrophysical background since their contribution to the X-ray emission is expected to be much smaller than that of the cluster core in the reference band under consideration of 6.0-8.0~keV surrounding the Fe-K lines which will strongly constrain the Doppler shifting and broadening.

\section{Results}\label{sec:results}

\subsection{Projected Velocity Fields}\label{sec:fullres_maps}

We first examine maps of X-ray surface brightness (in the 0.6-9.0~keV band), line shift, and line width at the full resolution of our simulations, along different lines of sight, presented in Figures \ref{fig:novisc_S_X}-\ref{fig:visc_sigma}. The line shift and line width have been computed by integrating the emission-weighted velocity field along the line of sight:

\begin{eqnarray}
\mu_\ell(\boldsymbol{\chi}) &=& \displaystyle\int{v_\ell}({\bf r})w_{\epsilon}({\bf r})\hat{\Bell}{\cdot}d{\bf r} \\
\sigma^2_\ell(\boldsymbol{\chi}) &=& \displaystyle\int{v_\ell^2}({\bf r})w_{\epsilon}({\bf r})\hat{\Bell}{\cdot}d{\bf r} - \mu_\ell^2(\boldsymbol{\chi}),
\end{eqnarray}

where

\begin{equation}
w_\epsilon({\bf r}) = \frac{\epsilon({\bf r})}{\displaystyle{\int{\epsilon({\bf r})}\hat{\Bell}{\cdot}d{\bf r}}},
\end{equation}

$\epsilon$ is the X-ray emissivity, $\boldsymbol{\chi}$ is the 2D coordinate in the plane of the sky, and ${\bf r}$ is the 3D coordinate in the simulation domain. In each of these figures, the projected position of the cluster potential minimum is marked with a cross symbol, and the simulated {\it Hitomi}/SXS pointing is indicated by a square. The surface brightness maps in Figures \ref{fig:novisc_S_X} and \ref{fig:visc_S_X} are shown to faciliate easy identification of features in the velocity maps with respect to the location of the cold fronts.

The line shift maps are shown in Figures \ref{fig:novisc_mu} and \ref{fig:visc_mu}. The gas regions underneath the cold fronts (where ``under'' and ``over'' refer to the directions closer and further away from the cluster core, respectively) surrounding the cluster center are regions which can be observed with a significant line shift, provided that the line of sight is not perpendicular to the merger plane, i.e.~along the $z$-axis. When viewed along the $z$-axis ($\theta = 0^\circ$), the line shift has a random pattern across the core and is very modest, with values $|\mu_{\ell}| \simlt 60$~km~s$^{-1}$ in the inviscid simulation, and even less in the viscous simulation. This is easily understood: this axis is perpendicular to the plane of the gas motions induced by the merger, and so although there are gas motions in this direction also, they are symmetric across the $x$-$y$ plane of the simulation domain and hence cancel each other out. Whatever smaller-scale turbulence may be driven in this direction has an average velocity of nearly zero also. 

As the line of sight is rotated from the $z$-axis to the $x$-axis, the magnitude of the line shift underneath the cold fronts increases, in keeping with the fact that our line of sight now includes components of the velocity field within the cluster merger plane. Underneath the southern cold front, with its edge roughly 100~kpc to the south of the core, $\mu_{\ell}$ increases to $\mu_{\ell} \sim 300$~km~s$^{-1}$ when viewed along the $x$-axis ($\theta = 90^\circ$). The velocity of the gas underneath northern cold front, roughly 200~kpc to the north of the core, has a line shift of $\mu_{\ell} \sim -250$~km~s$^{-1}$. Within the region of the simulated SXS pointing, there is a velocity gradient across the core (on opposite sides of the cluster potential minimum) of several hundred km~s$^{-1}$. All of these features, which are on length scales comparable to the size of the cold fronts themselves, are common to both the inviscid and the viscous simulations. Unsurprisingly, the inviscid simulation is more disturbed by instabilities and turbulence on smaller length scales than the viscous simulation.

The line width maps are shown in Figures \ref{fig:novisc_sigma} and \ref{fig:visc_sigma}. Both the core region and the northern cold front are regions with significant line broadening. This is true for both the inviscid and viscous simulations, though the line widths are somewhat larger in the inviscid case, with largest values of $\sigma \sim 200-250$~km~s$^{-1}$ in that simulation versus $\sigma \sim 150-200$~km~s$^{-1}$ in the viscous simulation. As in the case of the line shift, the major difference in the maps of line width between the two simulations is that the inviscid simulation shows more evidence of turbulence and instabilities than the viscous simulation, but the large-scale features are very similar.

The {\it Chandra} observations of the Perseus cluster have shown clear indications of sloshing gas motions, as evidence by spiral-shaped cold fronts. When viewing cold fronts along a line of sight nearly aligned with the plane of the gas motions, the associated surface brightness and temperature jumps are still visible but the spiral pattern is far less obvious (see the last panels of Figures \ref{fig:novisc_S_X} and \ref{fig:visc_S_X}). The {\it Hitomi} observations of the Perseus cluster reveal a velocity gradient across the cluster core, but such gradients would only be viewable in a sloshing cluster core if the line of sight is not perpendicular to the plane of those motions defined by the orbital plane of the main cluster and its perturber. To satisfy the twin conditions of viewing both spiral-shaped cold fronts and a gradient in the line shift from the same sloshing motions, we therefore suggest that the system must be viewed at an intermediate angle between the extremes of perpendicular to and parallel with this plane. For the rest of this work, we adopt the line of sight defined by $\theta = 45^\circ$ as a ``fiducial'' orientation which provides a match to these features of both the {\it Chandra} and {\it Hitomi} data. 

\subsection{Mock {\it Hitomi}/SXS Observations}\label{sec:mock_obs}

We make mock observations of our simulated clusters using the procedure described in Section \ref{sec:mock_obs_method}. In Figures \ref{fig:novisc_S_X}-\ref{fig:visc_sigma}, we previously noted the position of our simulated SXS pointing. This pointing was chosen to provide a qualitative match to location of the existing observation of the Perseus cluster and to capture the dynamics of the core region bounded by the innermost cold fronts. For all of our mock observations, our exposure time is 300~ks. For our simulated cluster, this exposure time gives counting statistics which are similar to those from the {\it Hitomi} observations of Perseus as detailed in H16 and H17. For each mock observation, we obtain the spectrum within each of 9 $1' \times 1'$ regions which tile the SXS field of view, in order to at least somewhat mitigate the effects of the PSF. We also obtain the spectrum within two larger regions, an ``Inner'' region and an ``Outer'' region (shown in Figure \ref{fig:compare_mu}), the former close to the cluster potential minimum and the latter somewhat further away, to measure the velocity difference between these regions. These regions are similar to the regions chosen in H16 (see their Figure 3). We fit each spectrum within the 6.0-8.0~keV band, in the region of the Fe-K lines, to a \code{tbabs*bapec} model with XSPEC\footnote{\url{http://heasarc.gsfc.nasa.gov/xanadu/xspec/}}, again assuming \citet{asp09} abundances. For each fit, we hold the Galactic hydrogen column and the metallicity parameters fixed at the input values noted above. All other parameters are free to vary. For the exposure time we simulated, the typical statistical 1$\sigma$ error on the line shift for the $1' \times 1'$ regions is $\sim \pm~10$~km~s$^{-1}$, and the typical statistical 1$\sigma$ error on the line width is $\sim \pm~10$~km~s$^{-1}$.

\subsubsection{Line Shift and Width Maps}\label{sec:sxs_maps}

Figure \ref{fig:compare_mu} shows maps of the line shift for both simulations, with the inviscid simulation in the left panels and the viscous simulation in the right panels, for our fiducial line of sight of $\theta = 45^\circ$. The full-resolution maps of X-ray surface brightness and line shift are also included in these panels to facilitate easy comparison. The ``Inner'' and ``Outer'' regions used to compute the velocity gradient across the core are marked, as well as the cluster potential minimum with a cross. Two maps of the line shift based on the mock SXS observations are shown: one with the $\sim$1' spatial PSF, and another with no PSF, but with the same spatial binning of the events into the 9 regions. 

The first thing to note about the plots is that the large-scale features of the line shift (on scales $\sim$1' and above) are accurately captured by the mock observations and the spectral fitting. In the northeast region of the SXS pointing, nearest the core region, the line shift is negative, while in the regions further away from the core, the line shift is positive. This is in accordance with the behavior of the line shift in the full-resolution maps. The extreme values of the line shift seen in the full-resolution maps are not present in the SXS-based maps, since the regions moving at these velocities make up a small portion of the gas emission, and at the lower resolution of SXS the line shift is dominated by gas motions with somewhat smaller values of the velocity magnitude. Because most of the differences in the line shift between the inviscid and viscous simulations occur at scales smaller than the SXS PSF, the line shift maps between the two different simulations at this resolution look very similar.

From these maps it can also be seen that the PSF has an effect on the estimated line shift. Photons emitted from the core region will be scattered into nearby regions, biasing the line shift in these regions in the direction of the line shift of the core. Depending on the brightness in a given 1' region, as many as $\sim$20-30\% of its photons will be scattered into it from a neighboring region. This effect was previously discussed in the context of {\it Hitomi} observations of galaxy clusters in H16, H17, Z16, and \citet{kit14}. The magnitude of this shift can be as much as $\sim 40-50$~km~s$^{-1}$ for $1' \times 1'$ regions near the core, larger than the statistical error for the same regions and comparable to the systematic error. 
  
Figure \ref{fig:compare_sigma} shows similar maps of the line width for both simulations, with the inviscid simulation in the left panels and the viscous simulation in the right panels, for our fiducial line of sight of $\theta = 45^\circ$, along with the corresponding full-resolution maps. Two of the main features from the line shift maps also manifest themselves in the line width maps: the large-scale features of the line shift (on scales $\sim$1' and above) are accurately captured by the mock observations and the spectral fitting, and the extreme values of the line shift seen in the full-resolution maps are not present in the SXS-based maps. The value of the line width in the different pixels is not biased to the same degree by the effect of the PSF as the line shift: the bias on the line width is typically $\sim 15-25$~km~s$^{-1}$, comparable to the 1$\sigma$ statistical error on the line width. Differences between the inviscid and viscous simulations are more apparent in the maps of the line width than they are of the line shift.

\subsubsection{Properties of the Velocity Field in the Inner and Outer Regions of the Core}\label{sec:velocity_gradient}

\begin{figure*}
  \begin{center}
  \includegraphics[width=0.48\textwidth]{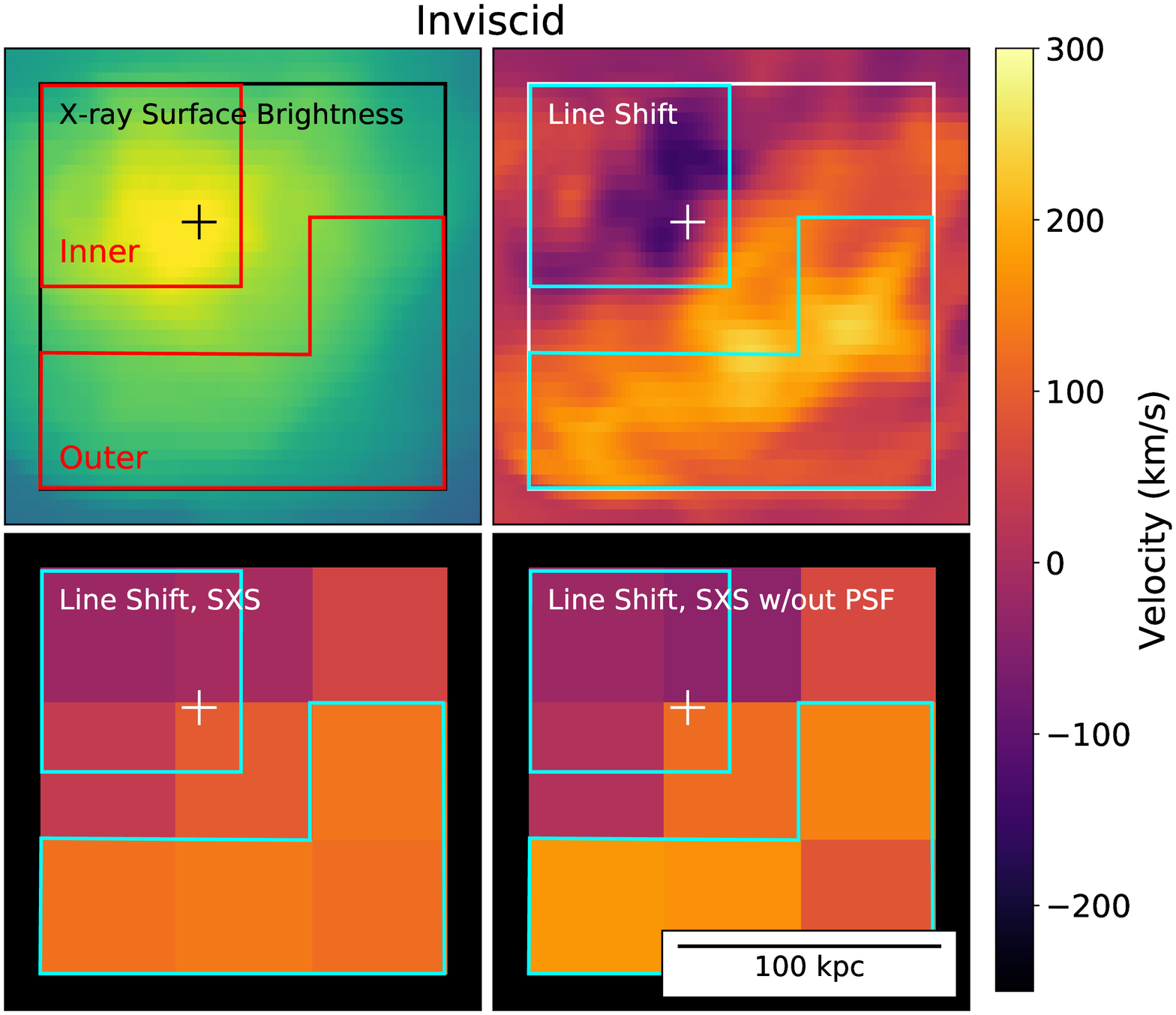}
  \includegraphics[width=0.48\textwidth]{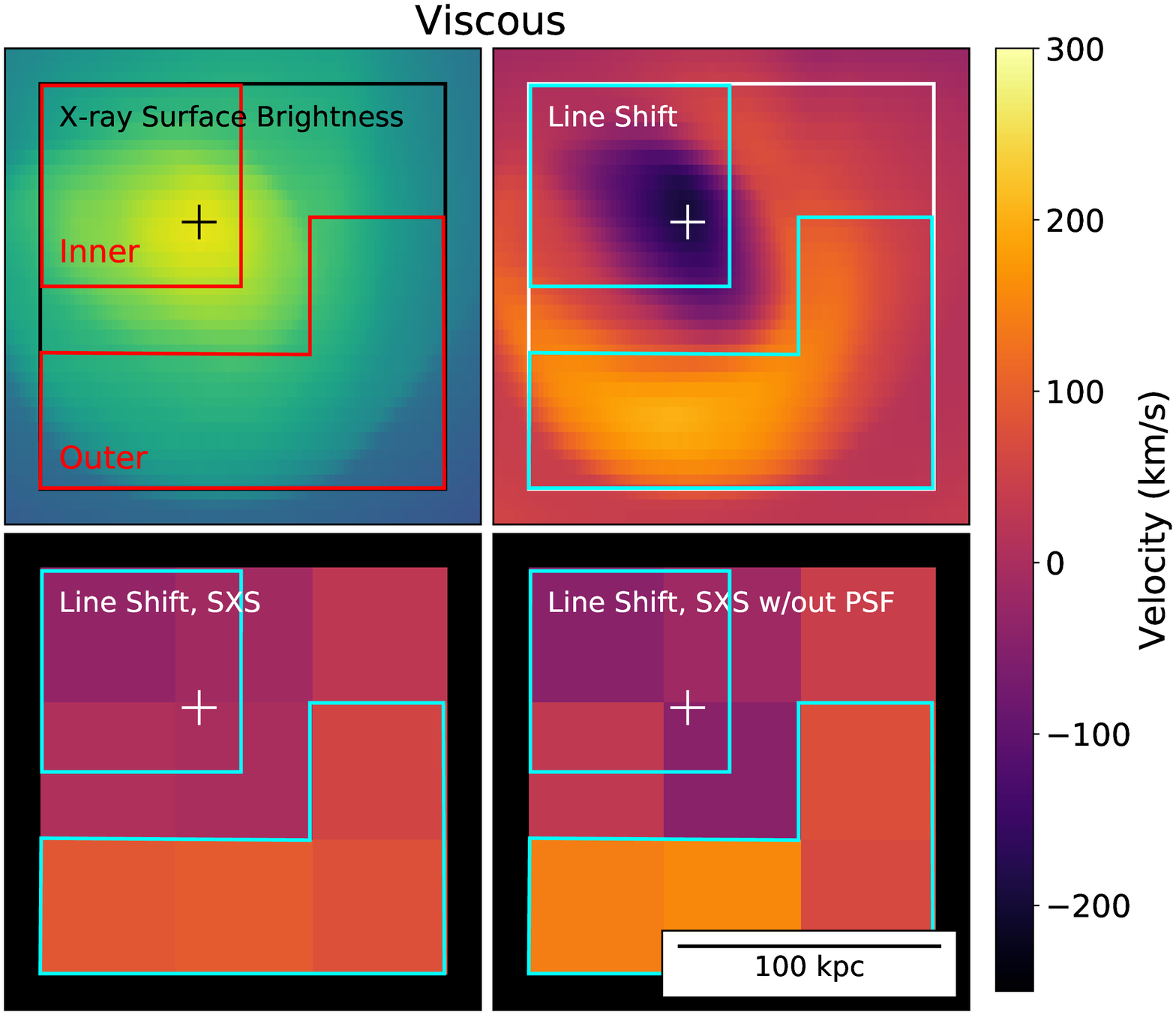}
  \caption{Comparisons of line shifts obtained from the simulation itself and estimated from spectral fitting to mock X-ray observations, for both simulations. In both sets of panels, the plotted quantities from the top-left going counterclockwise are: X-ray surface brightness from the simulation, line shift from the simulation, line shift measured by SXS, and line shift measured by SXS in the absence of spatial PSF effects. The black or white square in each sub-panel shows the SXS field of view. The ``Inner'' and ``Outer'' regions where the line shift is measured for Figures \ref{fig:shift_profiles} and \ref{fig:shear_profiles} are also marked. The cross indicates the position of the of the gravitational potential minimum of the cluster.\label{fig:compare_mu}}
  \end{center}
  \end{figure*}

\begin{figure*}
\begin{center}
\includegraphics[width=0.48\textwidth]{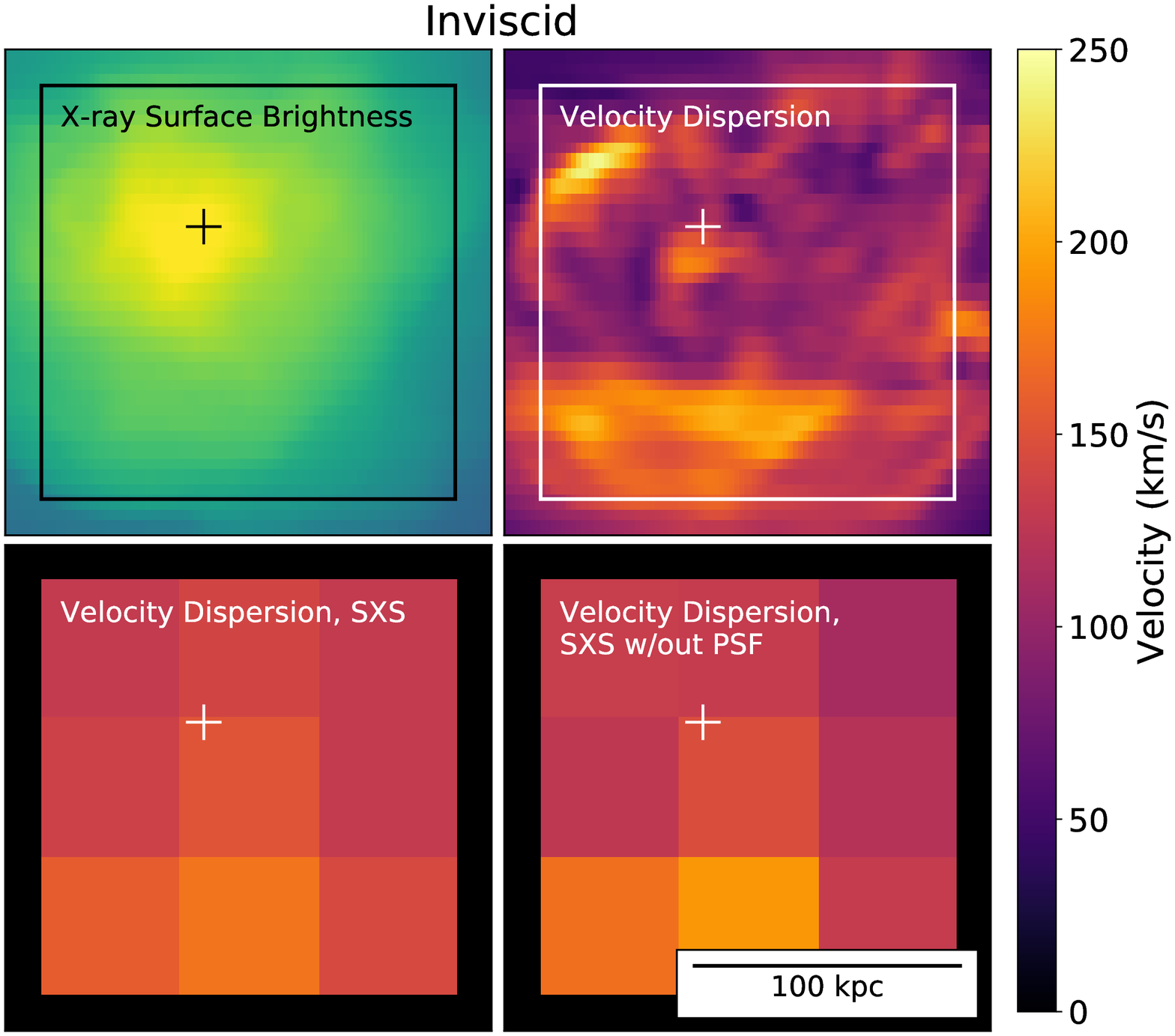}
\includegraphics[width=0.48\textwidth]{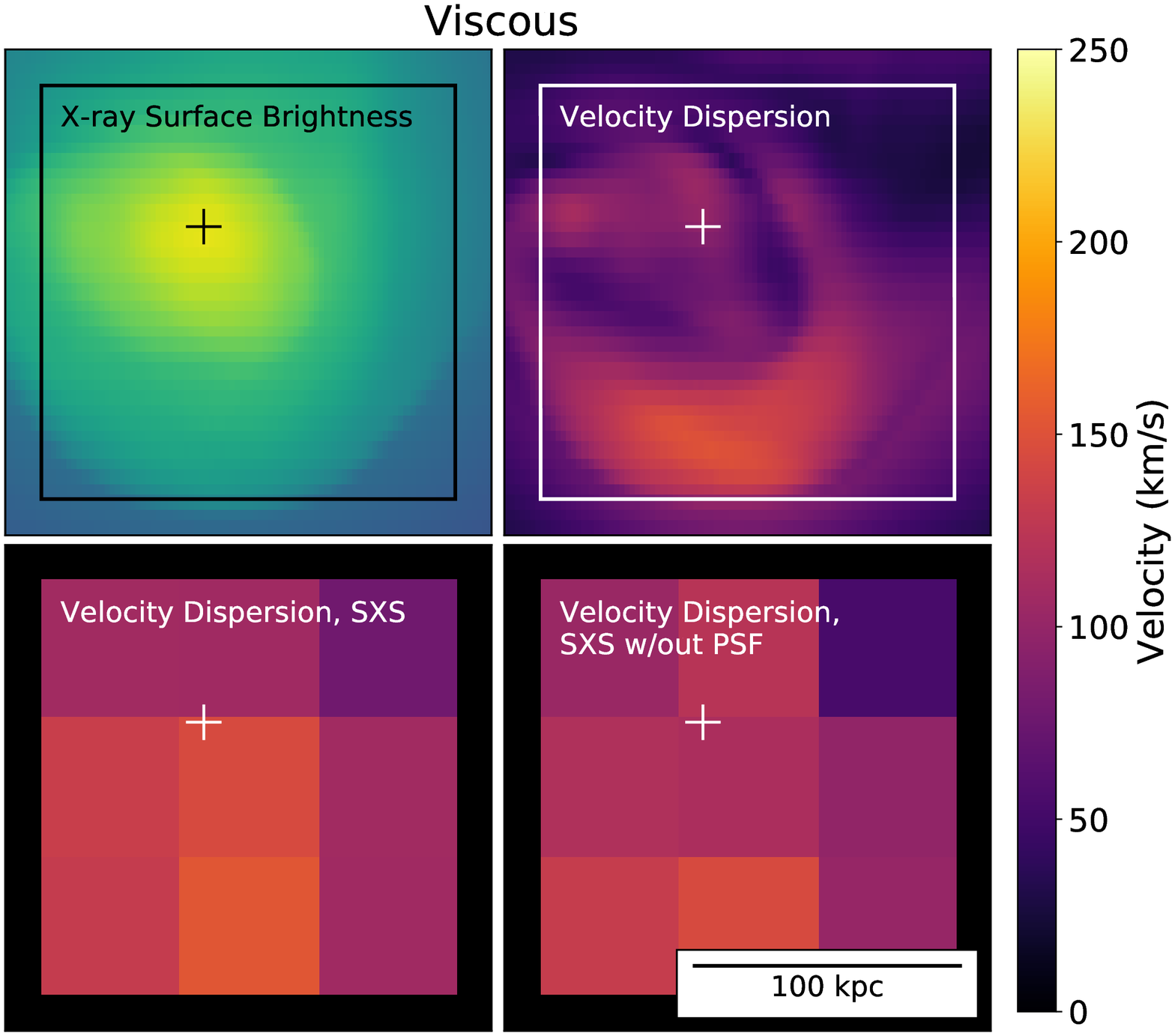}
\caption{Comparisons of line widths obtained from the simulation itself and estimated from spectral fitting to mock X-ray observations, for both simulations. In both sets of panels, the plotted quantities from the top-left going counterclockwise are: X-ray surface brightness from the simulation, line width from the simulation, line width measured by SXS, and line width measured by SXS in the absence of spatial PSF effects. The black or white square in each sub-panel shows the SXS field of view. The cross indicates the position of the of the gravitational potential minimum of the cluster.\label{fig:compare_sigma}}
\end{center}
\end{figure*}

\begin{figure*}
\begin{center}
\includegraphics[width=0.96\textwidth]{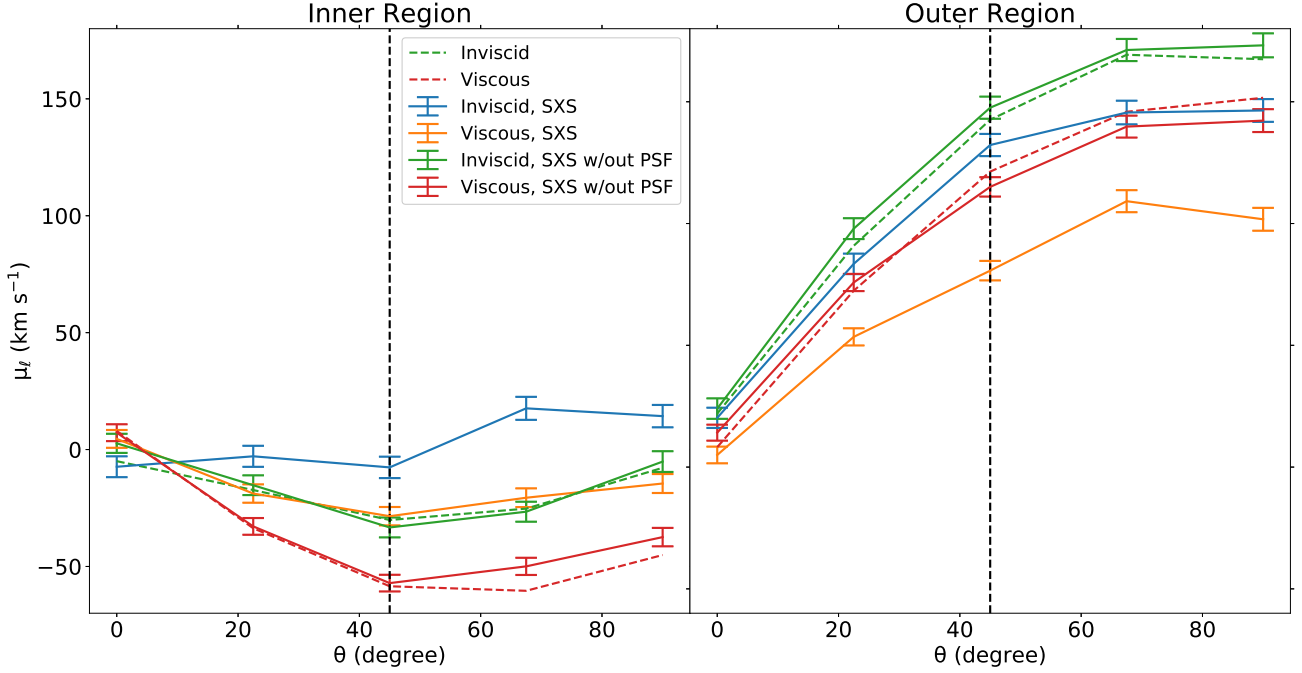}
\caption{Line shifts estimated from spectral fitting and obtained from the simulation. Left panel: The line shift from from the ``Inner'' region of the cluster as a function of viewing angle from both simulations, with and without the SXS PSF applied. Right panel: The line shift from the ``Outer'' region of the cluster as a function of viewing angle from both simulations, with and without the SXS PSF applied. The dashed vertical line in both panels indicates our fiducial orientation of $\theta = 45^\circ$.\label{fig:shift_profiles}}
\end{center}
\end{figure*}

We can use the ``Inner'' and ``Outer'' regions shown in Figure \ref{fig:compare_mu} to calculate line shifts in regions close to and far away from the cluster center and determine the velocity difference between these two regions, in a similar manner to what was done for Perseus in H16. The values of the line shift in these regions as a function of viewing angle are shown in Figure \ref{fig:shift_profiles}. For each of the two regions, the line shift in these regions begins near zero at $\theta = 0^\circ$, and its absolute value increases as $\theta$ approaches $90^\circ$, in line with the expectations from Section \ref{sec:fullres_maps}. Assuming the SXS PSF, the maximum difference in the line shift between the two models for viscosity is $\sim 50-60$~km~s$^{-1}$ in both regions (the blue and orange solid curves in both panels). With no PSF applied, the difference in the line shift between the two cases is slightly smaller, being $\sim 30-40$~km~s$^{-1}$ in both regions (the green and red solid curves in both panels). The green and red dashed curves in both panels show the line shift computed for the same regions by taking an average of the line shift from the full-resolution maps weighted by the surface brightness. These curves agree very well with the fitted values of the line shift from the mock observations without the PSF. This indicates that the difference in the line shift induced by the PSF from the ``true'' value is roughly $\sim 20-40$~km~s$^{-1}$, larger than the statistical errors on the line shift but smaller than or comparable to the error on the line shift from the gain uncertainty. 

\begin{figure}
\begin{center}
\includegraphics[width=0.49\textwidth]{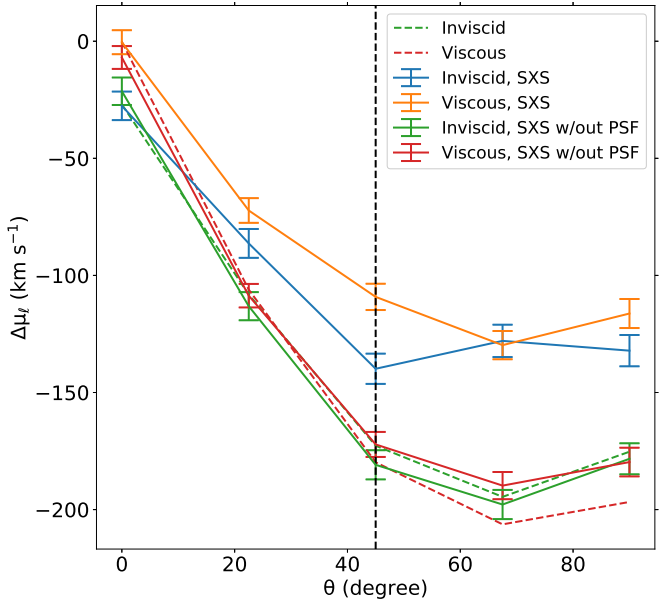}
\caption{Velocity gradient across the cluster core estimated from spectral fitting and obtained from the simulation, for both simulations and both PSF models, as a function of viewing angle. The dashed vertical line indicates our fiducial orientation of $\theta = 45^\circ$.\label{fig:shear_profiles}}
\end{center}
\end{figure}

Figure \ref{fig:shear_profiles} shows the velocity difference $\Delta{\mu_\ell}$ between these two regions as a function of viewing angle. The velocity difference between the two different simulations is roughly the same (regardless of whether or not the PSF is applied), indicating that viscosity has essentially no affect on this quantity. This is likely due to the fact that the velocity difference arises from gas motions that exist at length scales far above those which even a strong viscosity is able to damp. However, the difference between the mock observations with and without the application of the PSF is more dramatic--if the PSF is applied it decreases the velocity difference by  nearly $\sim 50-60$~km~s$^{-1}$ for $\theta > 45^\circ$. At our fiducial value of $\theta = 45^\circ$, $\Delta{\mu_\ell} \sim 110-130$~km~s$^{-1}$ if the SXS PSF is applied, and $\Delta{\mu_\ell} \sim 170-180$~km~s$^{-1}$ if it is not. The velocity gradient estimated directly from the full-resolution maps is given by the dashed lines, and is in agreement with the measurements with no PSF applied. This effect is easily understood to be due to the fact that the two regions are on opposite sides of the bright cluster core, and scattering of photons emitted from this region into the two different regions biases the line shift of each region slightly towards that of the core, reducing the difference between the two. 

\subsubsection{Velocity Dispersion}\label{sec:velocity_dispersion}

\begin{figure}
\begin{center}
\includegraphics[width=0.49\textwidth]{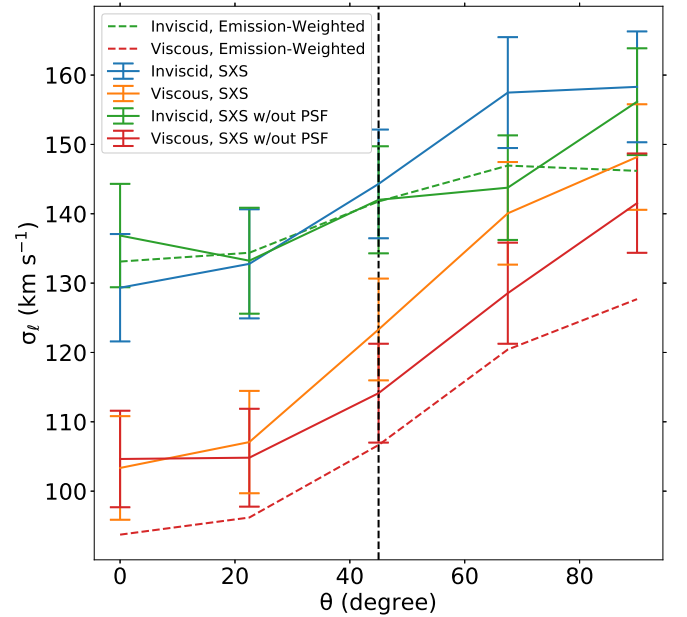}
\caption{Velocity dispersion in the core region as a function of viewing angle, where an average is taken over the 9 $1' \times 1'$ regions which tile the SXS field of view. The dashed vertical line indicates our fiducial orientation of $\theta = 45^\circ$.\label{fig:dispersion}}
\end{center}
\end{figure}

We also determined the velocity dispersion in the core region. Figure \ref{fig:dispersion} shows the average velocity dispersion in the core region as a function of viewing angle $\theta$, determined by fitting for the velocity dispersion in each of the 9 $1' \times 1'$ regions described above and averaging them. This was also done for the full resolution maps by taking the emission-weighted average within each of the 9 regions and averaging those, which are shown in Figure \ref{fig:dispersion} by the dashed lines. The PSF has little effect on the measurement of the velocity dispersion, which is expected, since effect was measured to be small in Section \ref{sec:sxs_maps}, and whatever differences exist are averaged out by taking the mean value over the entire core region. The viscous simulation has a velocity dispersion that measures roughly 20-30~km~s$^{-1}$ less than the inviscid simulation, regardless of viewing angle, a $\sim$~2-3$\sigma$ effect. This velocity dispersion is comparable to that measured for the Perseus cluster by {\it Hitomi}.

\begin{figure}
\begin{center}
\includegraphics[width=0.49\textwidth]{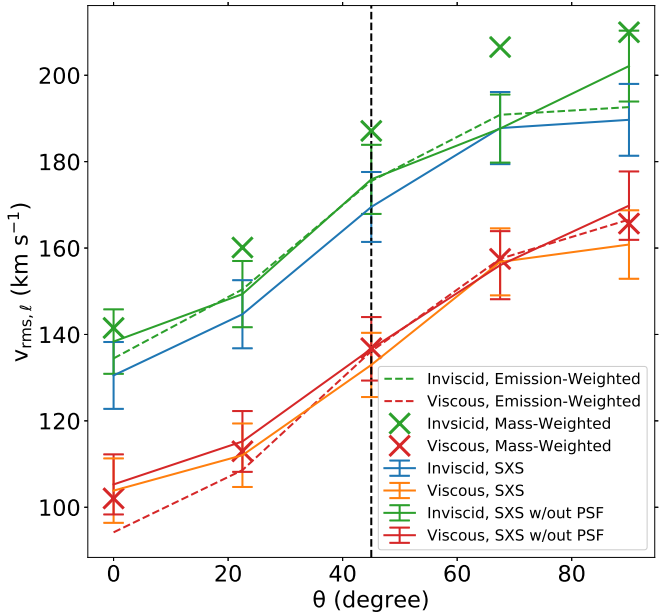}
\caption{Estimation of the total average velocity within the core as computed using Equation \ref{eqn:v_rms_obs} from the mock {\it Hitomi} observations, compared to data obtained from the simulation, as a function of viewing angle. The large ``$\times$'''s indicate the mass-weighted averaged velocity magnitude within the core region obtained from the simulation. The dashed vertical line indicates our fiducial orientation of $\theta = 45^\circ$.\label{fig:total_velocity}}
\end{center}
\end{figure}

\begin{figure}
\begin{center}
\includegraphics[width=0.49\textwidth]{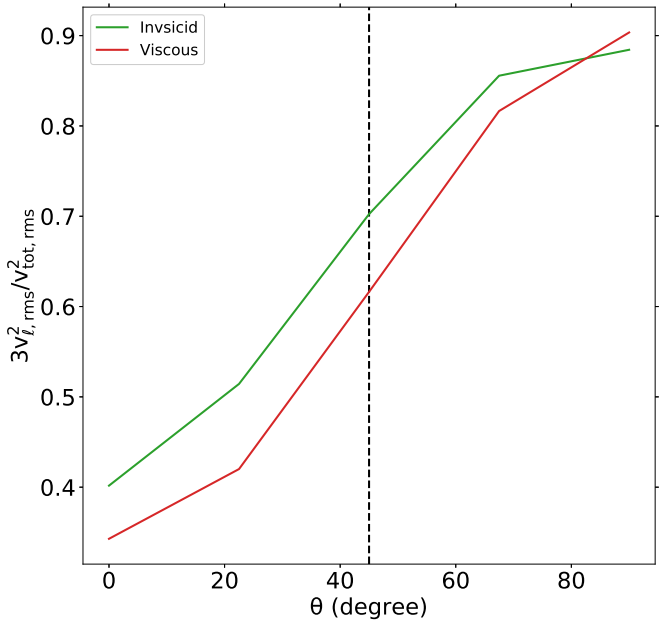}
\caption{The ratio of the estimated root-mean-squared velocity in the core, assuming isotropy, to the true root-mean-squared velocity, as a function of viewing angle $\theta$. The dashed vertical line indicates our fiducial orientation of $\theta = 45^\circ$.\label{fig:vtot_fraction}}
\end{center}
\end{figure}

\begin{figure*}
\begin{center}
\includegraphics[width=0.49\textwidth]{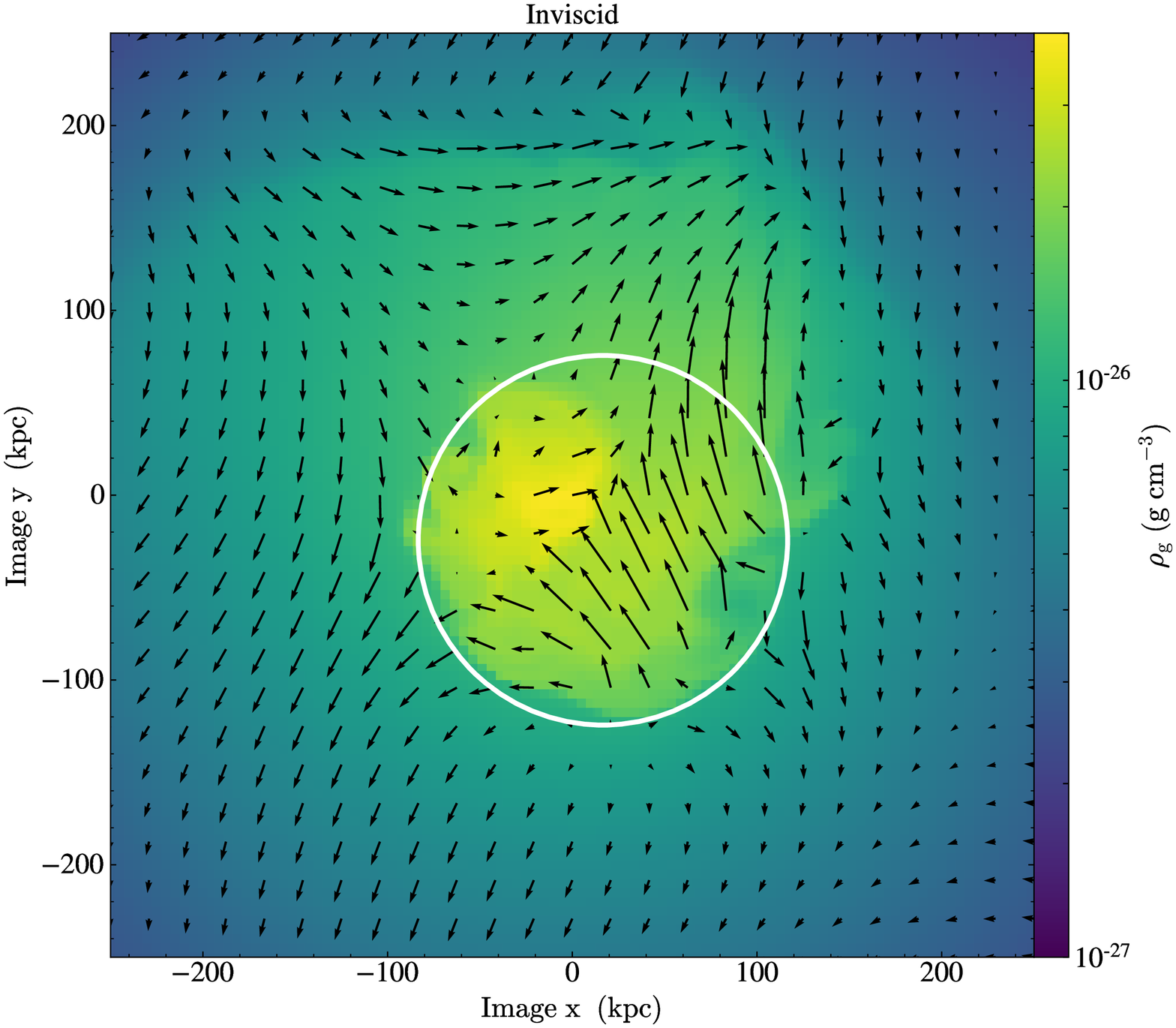}
\includegraphics[width=0.49\textwidth]{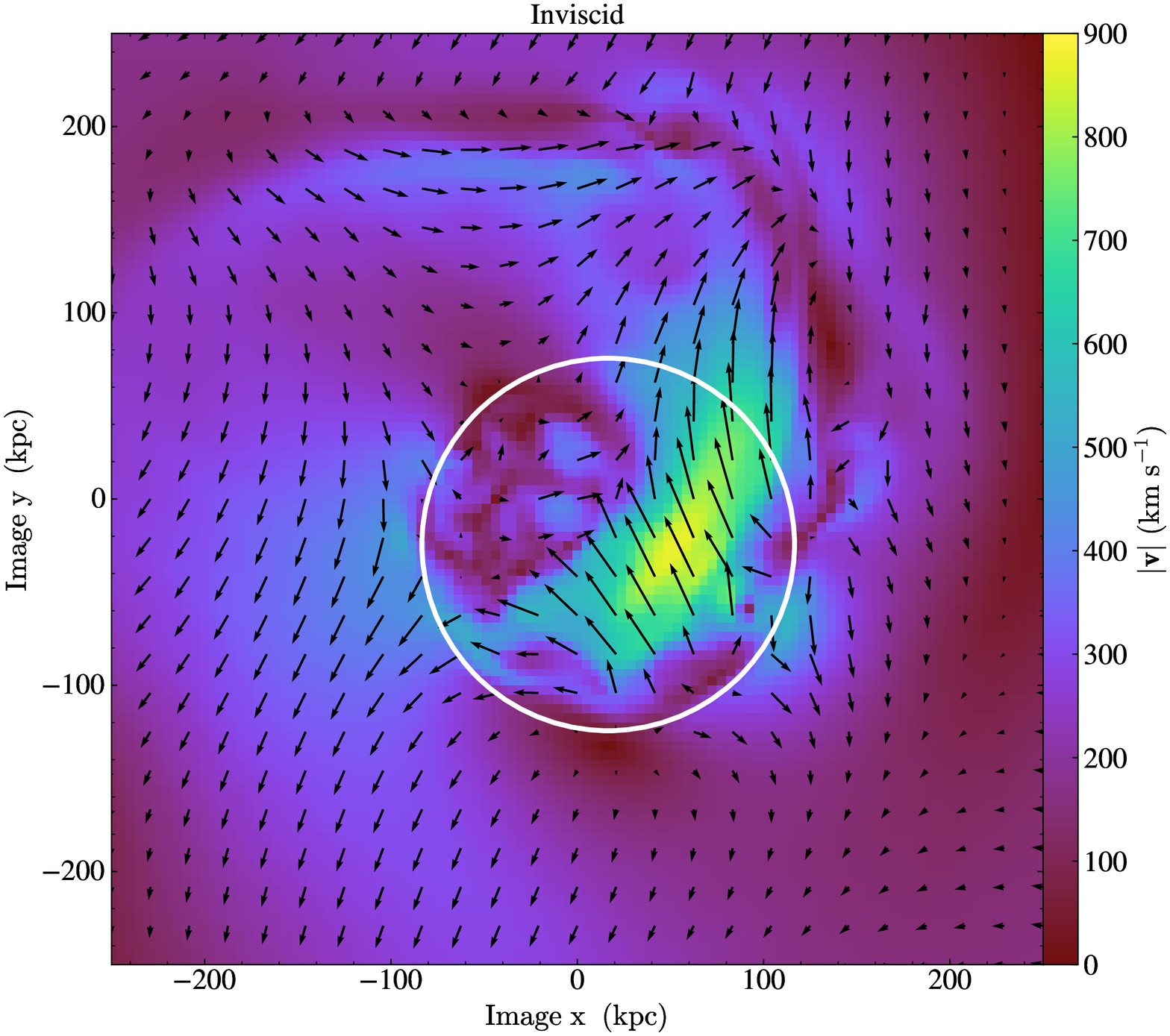}
\includegraphics[width=0.49\textwidth]{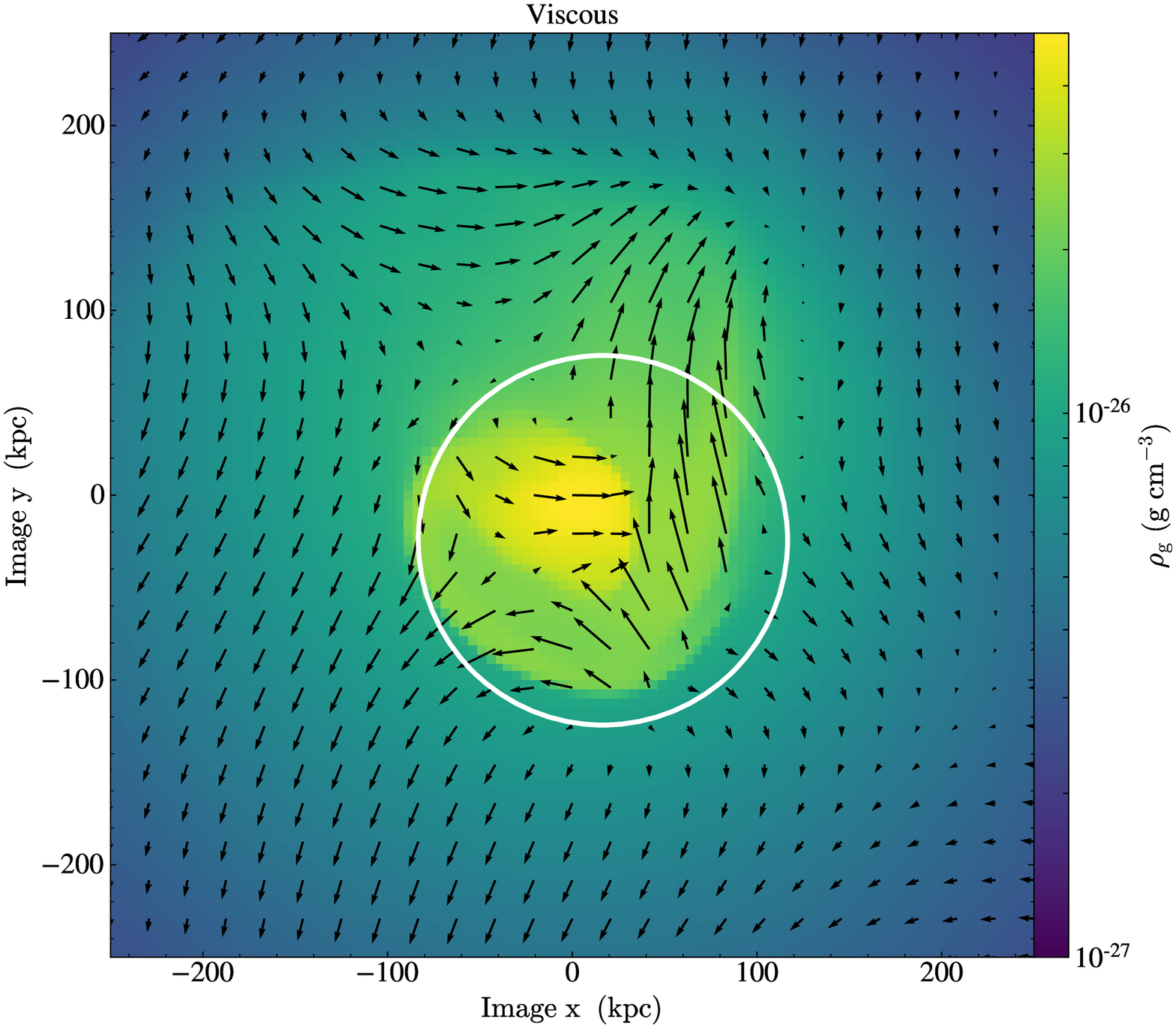}
\includegraphics[width=0.49\textwidth]{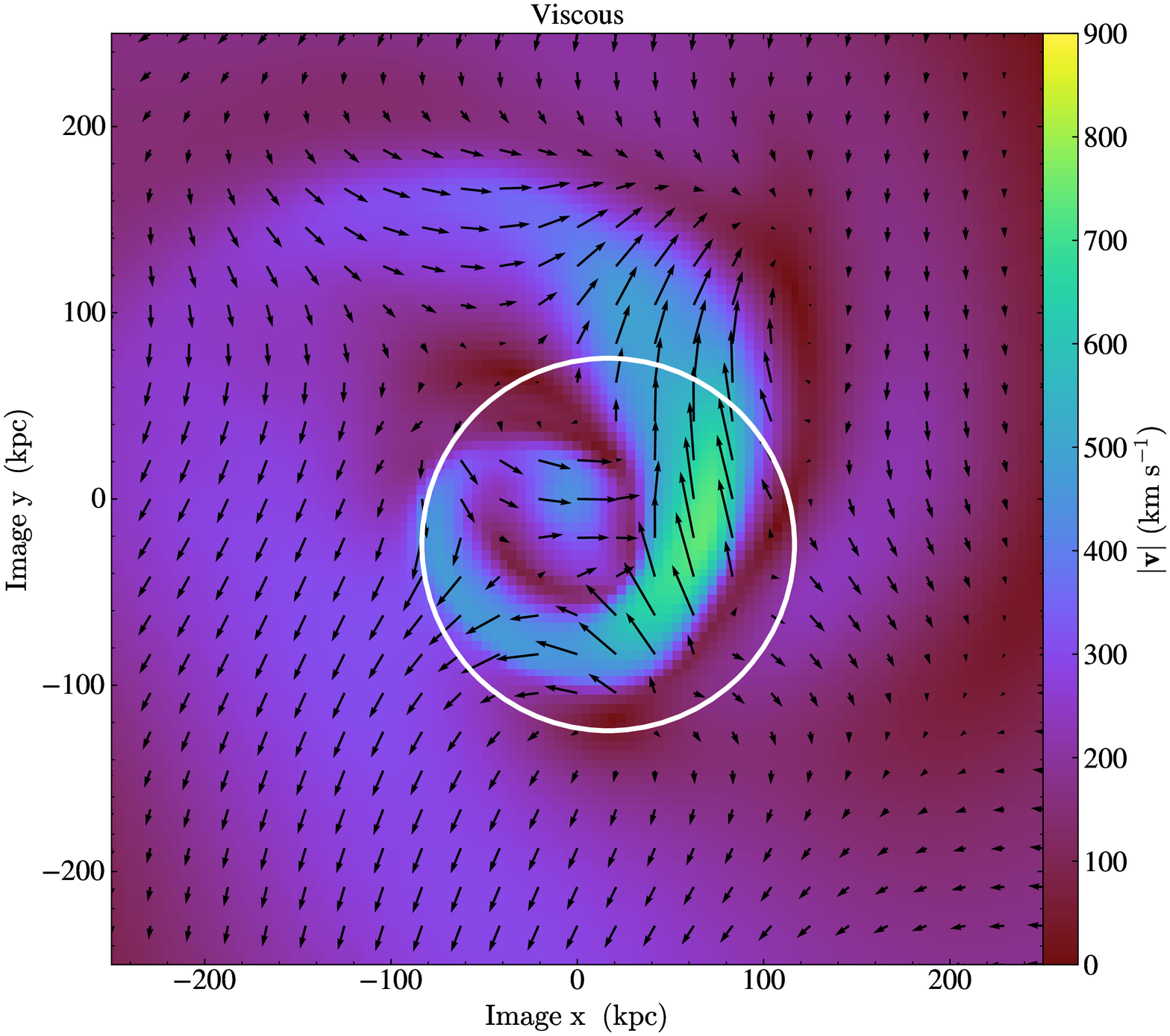}
\caption{Slices in the $x-y$ plane of density (left panels) and velocity magnitude (right panels) through the center of our cluster, for both the inviscid (top panels) and the viscous (bottom panels) simulations. Vectors indicate the direction of velocity within the plane. The white circle marks the region within which 3D velocity information is measured in Section \ref{sec:core_kinematics}. Each panel is 500~kpc on a side.\label{fig:slices}}
\end{center}
\end{figure*}

\subsubsection{Determining the Kinematic Properties of the Core}\label{sec:core_kinematics}

A primary goal of the {\it Hitomi} observations of Perseus (and of all similar future observations of clusters with microcalorimeters) is to determine the kinetic energy associated with gas motions. This requires summing the total contribution to the velocity field as measured from both the line shift and line width. However, knowing the velocity in the components perpendicular to our sight line is also required, which of course is unmeasurable.

We may estimate the average velocity in the core along our line of sight via the root-mean-squared velocity:

\begin{equation}
v_{\ell, \rm rms} = \sqrt{\langle{v_{\ell}^2}\rangle} = \sqrt{\langle\sigma_{\ell}^2+\mu_{\ell}^2\rangle}
\end{equation}

where the average is emission-weighted and is taken over the entire SXS field of view. Figure \ref{fig:total_velocity} shows the root-mean square velocity averaged over the entire core region, as a function of the line-of-sight angle $\theta$. Like the velocity dispersion, the SXS PSF has little effect on this measurement, but the inviscid and viscous cases are separated by approximately 40-50~km~s$^{-1}$, a roughly 2-3$\sigma$ difference. The total velocity estimated in the same way from the full-resolution maps (the dashed lines) is in good agreement with the simulated {\it Hitomi} measurements. For this plot, we also show the mass-weighted velocity component along the sight line averaged within a sphere of 100~kpc centered on the cluster potential minimum, given by the ``$\times$'' symbols in the figure, which also agrees well with the other measurements. This region is roughly the size covered by our simulated SXS pointing. 

In order to use this measurement to make an estimate of the velocity in the core averaged over all components, the simplest assumption to make is that the velocity field is isotropic. This implies:

\begin{equation}\label{eqn:v_rms_obs}
v_{\rm tot, rms} \sim \sqrt{3}v_{\ell, \rm rms}
\end{equation}

How accurate is the assumption of isotropy? Assuming the velocity field in the cluster core is at least somewhat anisotropic, the answer depends on the line of sight. The result is shown in Figure \ref{fig:vtot_fraction}, which shows the ratio of the estimate of the ``total'' root-mean-squared velocity in the core assuming isotropy and the true root-mean-squared velocity summed over all components as a function of our line of sight $\theta$. For our simulated cluster, if the cluster is being viewed along a sight line near the plane of the gas motions ($\theta = 90^\circ$), then assuming isotropy will provide an underestimate of the kinetic energy in the core by only $\sim10\%$. For the same reasons, assuming isotropy when observing the cluster along a line of sight perpendicular to this plane ($\theta = 0^\circ$) will underestimate the kinetic energy of the core by $\sim$60\%. For our fiducial line of sight of ($\theta = 45^\circ$), the kinetic energy is underestimated by $\sim$30\%. These results hold regardless of the viscosity of the simulation. These arguments demonstrate that if one is observing the spiral pattern of the cold fronts as we do in Perseus, any estimate of the total velocity in the core that depends on the line of sight velocity and assumes isotropy of the velocity components is likely to underestimate the kinetic energy in the core, at least if it is dominated by the sloshing motions. 

This is also illustrated in Figure \ref{fig:slices}, which shows slices through the density and velocity magnitude through the $x-y$ plane of both simulations at our chosen epoch, with velocity vectors overlaid. A white circle marks the region within which the mass-weighted velocities were computed for Figure \ref{fig:total_velocity}. One can see that there is a substantial flow of velocity in the $x-y$ plane, and in fact the fastest flow in this region is mainly in the $y$-direction, up to 800-900~km~s$^{-1}$, which is not covered by any of the lines of sight we simulated. If we measure from the simulation what the relative contributions to the kinetic energy are from the three principal components of the velocity within the central spherical region of 100~kpc, we find:

\begin{eqnarray}
\langle{v_x^2}\rangle/\langle{v_{\rm tot}^2}\rangle &\approx& 0.29, \\ 
\langle{v_y^2}\rangle/\langle{v_{\rm tot}^2}\rangle &\approx& 0.57, \\
\langle{v_z^2}\rangle/\langle{v_{\rm tot}^2}\rangle &\approx& 0.14, 
\end{eqnarray}

showing that the contribution from the $y$-component of the velocity, not visible from any of our sight lines, makes up nearly $\sim 60\%$ of the kinetic energy within this region.

This flow also is in gas that is less dense than the densest parts of the core by a factor of several, so its surface brightness would be less by over an order of magnitude, and thus not contribute as much to the emission-weighted velocity even if it were within our sight line. This indicates that there may be even faster flows in Perseus that are simply not seen because they are in fainter regions. 

Despite this, we find that the total kinetic energy in the core is still a small fraction of the thermal energy: $\approx$9.7\% in the inviscid simulation and $\approx$5.9\% in the viscous simulation. This amount of kinetic energy justifies calling the Perseus core ``quiescent'' in this qualified sense. Interestingly, this is true regardless of whether or not the ICM is viscous.

\section{Summary}\label{sec:summary}

In this work, we have examined the velocity field of the ICM of a simulated galaxy cluster similar in character to the Perseus cluster, in the sense that our cluster also possesses subsonic sloshing gas motions as evidenced by spiral-shaped cold fronts. We produced synthetic {\it Hitomi}/SXS observations of this velocity field and performed similar analyses to those performed on the Perseus data. We have reached the following conclusions:

\begin{itemize}
\item We find that sloshing motions can produce line shifts and widths comparable to that found by {\it Hitomi} in the Perseus cluster. Assuming our fiducial line of sight, we measure a velocity gradient between the inner and outer regions of the cluster core of roughly $\Delta{\mu_\ell} \sim 110-130$~km~s$^{-1}$ if the SXS PSF is modeled, and roughly $\Delta{\mu_\ell} \sim 170-180$~km~s$^{-1}$ if the PSF is assumed to be a delta function. The magnitude of the velocity gradient is not strongly affected by viscosity. Along the same sight line, we measure an average velocity dispersion of $\sigma \sim 150$~km~s$^{-1}$ within the core region for the inviscid simulation, and $\sigma \sim 110$~km~s$^{-1}$ for the viscous simulation.
\item We find that the $\sim$1'~PSF of the SXS has a non-negligible effect on the measurement of the line shift on opposite sides of the cluster core, due to the fact that the line shift in these regions is biased by the presence of photons scattering into them from the core region. We find that this effect of PSF scattering induces a bias of roughly $\sim 20-40$~km~s$^{-1}$ on the line shift in the direction of the line shift of the bright cluster core for the regions nearer to and farther away from the cluster core. The bias on the line shift for individual resolution elements may be even larger. The effect of this bias is to decrease the velocity gradient $\Delta{\mu_\ell}$ from its true value by $\sim 50-60$~km~s$^{-1}$. We find a less significant effect of the PSF on velocity broadening, provided that the velocity broadening is measured on spatial scales at and below the scale of the PSF. When the PSF is not applied, we find that the velocities estimated from spectral fitting are in agreement with those directly measured from the simulation.
\item If both the spiral-shaped cold fronts and the observed velocity gradient in the Perseus cluster core are due largely to the bulk motions induced by gas sloshing, this indicates that the system may be viewed along a line of sight that is somewhat inclined with respect to the midplane of the cold fronts. We find that an angle of $\sim$45$^\circ$ provides an orientation which shows both spiral cold fronts and a significant line shift difference between the inner and outer parts of the core.
\item We show that overall the line shifts and widths measured within the core region at the spatial resolution of the SXS are very similar between an inviscid simulation and a highly viscous simulation. Line widths are more affected by viscosity than velocity gradients across the core, but the difference in line width between the two simulations is not drastic. This indicates that the observed evidence of gas motions in the Perseus cluster are more than likely to originate from velocity structures with characteristic scales larger than those that even a strong viscosity is expected to damp. These scales are also near or larger than the SXS spatial resolution. This is expected and is in line with previous results \citep[][Z16]{ino03}. This is also not inconsistent with the interpretation given by H17 that the Gaussianity of the line shapes implies that the driving scale of turbulence is just below $\sim$100~kpc, since this would be roughly the scale of the sloshing motions themselves, and likely still larger than the dissipation scale of the turbulent motions. We also note that Z16 reported that bulk motions such as sloshing can also produce line shapes that are remarkably consistent with Gaussianity, even in the absence of a developed turbulent cascade. These considerations imply that drawing strong conclusions about the microphysics of the cluster plasma (in particular its viscosity) from the {\it Hitomi} results should be avoided, since they simply do not have the spatial resolution required to probe the scales where such effects would be manifest in the observations. For this, the resolution of {\it Athena}\footnote{\url{http://www.the-athena-x-ray-observatory.eu/}} or {\it Lynx}\footnote{\url{http://wwwastro.msfc.nasa.gov/lynx/}} is likely to be required. The fact that the viscosity of the ICM is likely to be at least an order of magnitude less than the Spitzer value (based on plasma physics considerations, see Section \ref{sec:hydro_sims}) strengthens this conclusion.
\item For these simulations, we find that an estimate of the average velocity within the core made from the mock {\it Hitomi} observations agrees well with the average velocity in the core obtained directly from the simulation. Yet, this does not rule out the possibility of gas motions which are much faster in portions of the core which are either less dense/bright or not within the line of sight. In our simulation, a substantial portion of the kinetic energy ($\sim$60\%) is in a velocity component that was not within the sight lines that we simulated. For Perseus, this effect could be particularly significant if our line of sight is not close to the plane of the sloshing motions. 
\item The kinetic energy in the core region measured from our simulation is still less than 10\% of the thermal energy, in agreement with the estimates made from the {\it Hitomi} observations of Perseus. This is true irrespective whether or not the ICM is viscous, indicating the reason for the ``quiescent'' nature of the plasma is the lack of strong drivers of gas motions in the core (such as a recent major merger) and not viscosity. Though the core of the Perseus cluster is likely quiescent in this sense, given the possibility that a significant velocity component may not be within our line of sight, this claim should be made with caution.
\end{itemize}

Other recent comparisons have been made between simulations and the {\it Hitomi} observations. \citet[][]{lau17} performed an analysis of mock {\it Hitomi} observations of clusters from cosmological simulations and isolated clusters with AGN feedback. They concluded that cosmic accretion and mergers could produce line-of-sight velocity dispersions and line shifts compatible with the {\it Hitomi} observations of Perseus, while AGN feedback is able to produce velocity dispersion measurements which are compatible with the {\it Hitomi} observations, but not the core-scale velocity gradient, since the turbulence driven by AGN feedback is too stochastic. They argued from these results that cosmic accretion/mergers and AGN feedback are complementary drivers of the velocity field in cluster cores, and their combination likely explains the level of shear and velocity dispersion seen in the Perseus cluster. 

A similar conclusion was reached by \citet{bou17}, who simulated AGN-driven jets and mock {\it Hitomi} observations in isolated galaxy cluster models, and models which included realistic cosmic substructure. Their isolated-cluster simulations show that though the AGN feedback drives turbulence, it is mostly confined to the jet lobe regions, and no significant line shift gradients are produced. By contrast, their simulations which include gas motions driven by substructure produce line shift gradients and turbulence throughout the core region which are comparable to those observed in Perseus. They also concluded that a combination of AGN feedback and cosmic accretion is likely necessary to produce the gas motions which are inferred by the {\it Hitomi} observations. We also note that \citet{hil17} used simulations of jet-driven AGN feedback to show that velocity dispersions comparable to that observed by {\it Hitomi} could be produced.

By contrast, we conclude that sloshing motions driven by the ``cosmic accretion'' of a smaller subcluster alone are sufficient to produce both the signatures of core-scale velocity gradients and velocity dispersion. This is true even if the viscosity is unrealistically high, since the dominant contribution to the line width arises from gas motions with characteristic scales greater than the dissipation scale. Needless to say, our simulations do not include the effects of cooling or AGN feedback, which are essential for a more complete model of the dynamics of the core of Perseus. In particular, if our simulations had the effects of AGN feedback they would likely show a stronger velocity dispersion near the AGN, as reported by H17. It should also be noted in this context that since the AGN plays an outsized role in the dynamics of the cluster core in Perseus, that the sloshing motions themselves may be the product not of cosmic accretion but of the AGN feedback. \citet{fab11} argued that ICM structures seen in Perseus within the central $\sim$100~kpc of the cluster center are due to AGN activity, whereas structures outside of this region are due to cosmic accretion and mergers. Given that cool-core clusters such as Perseus have steep entropy profiles, their atmospheres are highly stratified and spiral structures in the velocity field can develop naturally by any process that produces an offset between gas and dark matter in the core, provided there is enough angular momentum imparted to the gas. Investigating all of these possibilities is left for future work.

\acknowledgments
We thank the anonymous referee whose comments improved this paper. This work required the use and integration of a number of software packages for astronomy:
\begin{enumerate}
\item{yt \citep{tur11}\footnote{\url{http://yt-project.org}}}
\item{pyXSIM \citep{zuh14}}
\item{SOXS}
\item{AstroPy \citep{ast13}\footnote{\url{http://www.astropy.org}}}
\item{APLpy\footnote{\code{\url{http://aplpy.github.io}}}}
\item{AstroPy Regions\footnote{\url{http://astropy-regions.readthedocs.io}}}
\item{XSPEC}
\item{CIAO\footnote{\url{http://cxc.harvard.edu/ciao/}}}\newline
\end{enumerate} 
We are thankful to the developers of these packages. The authors thank Mark Bautz for useful discussions. JAZ acknowledges support through Chandra Award Number G04-15088X issued by the Chandra X-ray Center, which is operated by the Smithsonian Astrophysical Observatory for and on behalf of NASA under contract NAS8-03060. EDM and EB acknowledge support from NASA grant NNX15AC76G. Calculations were performed using the computational resources of the Advanced Supercomputing Division at NASA/Ames Research Center.


\begin{thebibliography}{}
\bibitem[Ascasibar \& Markevitch(2006)]{AM06} Ascasibar, Y., \& Markevitch, M. 2006, \apj, 650, 102
\bibitem[Asplund et al.(2009)]{asp09} Asplund, M., Grevesse, N., Sauval, A.~J., \& Scott, P.\ 2009, \araa, 47, 481 
\bibitem[Astropy Collaboration et al.(2013)]{ast13} Astropy Collaboration, Robitaille, T.~P., Tollerud, E.~J., et al.\ 2013, \aap, 558, A33
\bibitem[Banerjee \& Sharma(2014)]{ban14} Banerjee, N., \& Sharma, P.\ 2014, \mnras, 443, 687
\bibitem[Bartalucci et al.(2017)]{bar17} Bartalucci, I., Arnaud, M., Pratt, G.~W., et al.\ 2017, \aap, 598, A61 
\bibitem[Bautz et al.(2009)]{bau09} Bautz, M.~W., Miller, E.~D., Sanders, J.~S., et al.\ 2009, \pasj, 61, 1117
\bibitem[Biffi et al.(2012)]{bif12} Biffi, V., Dolag, K., B{\"o}hringer, H., \& Lemson, G.\ 2012, \mnras, 420, 3545
\bibitem[Biffi et al.(2013)]{bif13} Biffi, V., Dolag, K., B{\"o}hringer, H.\ 2013, \mnras, 428, 1395 
\bibitem[Boehringer et al.(1993)]{boe93} Boehringer, H., Voges, W., Fabian, A.~C., Edge, A.~C., \& Neumann, D.~M.\ 1993, \mnras, 264, L25 
\bibitem[Bourne \& Sijacki(2017)]{bou17} Bourne, M.~A., \& Sijacki, D.\ 2017, arXiv:1705.07900 
\bibitem[Braginskii(1965)]{bra65} Braginskii, S.~I.\ 1965, Reviews of Plasma Physics, 1, 205
\bibitem[Brunetti \& Lazarian(2007)]{bru07} Brunetti, G., \& Lazarian, A.\ 2007, \mnras, 378, 245
\bibitem[Bulbul et al.(2012)]{bul12} Bulbul, G.~E., Smith, R.~K., Foster, A., et al.\ 2012, \apj, 747, 32 
\bibitem[Churazov et al.(2000)]{chu00} Churazov, E., Forman, W., Jones, C., \& B{\"o}hringer, H.\ 2000, \aap, 356, 788 
\bibitem[Churazov et al.(2003)]{chu03} Churazov, E., Forman, W., Jones, C., B{\"o}hringer, H.\ 2003, \apj, 590, 225
\bibitem[Churazov et al.(2004)]{chu04} Churazov, E., Forman, W., Jones, C., Sunyaev, R., B\"{o}hringer, H.\ 2004, \mnras, 347, 29
\bibitem[Churazov et al.(2012)]{chu12} Churazov, E., Vikhlinin, A., Zhuravleva, I., et al.\ 2012, \mnras, 421, 1123
\bibitem[Colella \& Woodward(1984)]{col84} Colella, P., \& Woodward, P.~R.\ 1984, Journal of Computational Physics, 54, 174
\bibitem[Dennis \& Chandran(2005)]{den05} Dennis, T.~J., \& Chandran, B.~D.~G.\ 2005, \apj, 622, 205
\bibitem[de Plaa et al.(2012)]{dep12} de Plaa, J., Zhuravleva, I., Werner, N., et al.\ 2012, \aap, 539, A34
\bibitem[Donnert et al.(2013)]{don13} Donnert, J., Dolag, K., Brunetti, G., \& Cassano, R.\ 2013, \mnras, 429, 3564
\bibitem[{Dubey} et~al.(2009)]{dub09} {Dubey}, A., {Antypas}, K., {Ganapathy}, M.~K., {Reid}, L.~B., {Riley}, K.~M., {Sheeler}, D., {Siegel}, A., {Weide}, K. Extensible component based architecture for FLASH, a massively parallel, multiphysics simulation code. Parallel Computing 35~(10-11), 512--522.
\bibitem[Eckert et al.(2013)]{eck13} Eckert, D., Molendi, S., Vazza, F., Ettori, S., \& Paltani, S.\ 2013, \aap, 551, A22 
\bibitem[Evrard et al.(1996)]{evr96} Evrard, A.~E., Metzler, C.~A., \& Navarro, J.~F.\ 1996, \apj, 469, 494
\bibitem[Ezer et al.(2017)]{eze17} Ezer, C., Bulbul, E., Nihal Ercan, E., et al.\ 2017, \apj, 836, 110 
\bibitem[Fabian et al.(2000)]{fab00} Fabian, A.~C., Sanders, J.~S., Ettori, S., et al.\ 2000, \mnras, 318, L65 
\bibitem[Fabian et al.(2002)]{fab02} Fabian, A.~C., Celotti, A., Blundell, K.~M., Kassim, N.~E., \& Perley, R.~A.\ 2002, \mnras, 331, 369 
\bibitem[Fabian et al.(2003)]{fab03} Fabian, A.~C., Sanders, J.~S., Crawford, C.~S., et al.\ 2003, \mnras, 344, L48
\bibitem[Fabian et al.(2006)]{fab06} Fabian, A.~C., Sanders, J.~S., Taylor, G.~B., et al.\ 2006, \mnras, 366, 417 
\bibitem[Fabian et al.(2011)]{fab11} Fabian, A.~C., Sanders, J.~S., Allen, S.~W., et al.\ 2011, \mnras, 418, 2154 
\bibitem[Foster et al.(2012)]{fos12} Foster, A.~R., Ji, L., Smith, R.~K., \& Brickhouse, N.~S.\ 2012, \apj, 756, 128 
\bibitem[Fujita et al.(2004)]{fuj04} Fujita, Y., Matsumoto, T., \& Wada, K.\ 2004, \apjl, 612, L9
\bibitem[Fujita et al.(2015)]{fuj15} Fujita, Y., Takizawa, M., Yamazaki, R., Akamatsu, H., \& Ohno, H.\ 2015, \apj, 815, 116 
\bibitem[Hillel \& Soker(2017)]{hil17} Hillel, S., \& Soker, N.\ 2017, \mnras, 466, L39 
\bibitem[Hitomi Collaboration et al.(2016)]{hit16} Hitomi Collaboration, Aharonian, F., Akamatsu, H., et al.\ 2016, \nat, 535, 117 (H16)
\bibitem[Hitomi Collaboration et al.(2017a)]{hit17a} Hitomi Collaboration, Aharonian, F., Akamatsu, H., et al.\ 2017, arXiv:1711.00240 (H17)
\bibitem[Hitomi Collaboration et al.(2017b)]{hit17b} Hitomi Collaboration, Aharonian, F., Akamatsu, H., et al.\ 2017, arXiv:1710.04648 
\bibitem[Inogamov \& Sunyaev(2003)]{ino03} Inogamov, N.~A., \& Sunyaev, R.~A.\ 2003, Astronomy Letters, 29, 791
\bibitem[Khatri \& Gaspari(2016)]{kha16} Khatri, R., \& Gaspari, M.\ 2016, \mnras, 463, 655 
\bibitem[Kitayama et al.(2014)]{kit14} Kitayama, T., Bautz, M., Markevitch, M., et al.\ 2014, arXiv:1412.1176 
\bibitem[Kolmogorov(1941)]{kol41} Kolmogorov, A.\ 1941, Akademiia Nauk SSSR Doklady, 30, 301
\bibitem[Kunz et al.(2014)]{kun14} Kunz, M.~W., Schekochihin, A.~A., \& Stone, J.~M.\ 2014, Physical Review Letters, 112, 205003
\bibitem[Lau et al.(2017)]{lau17} Lau, E.~T., Gaspari, M., Nagai, D., \& Coppi, P.\ 2017, arXiv:1705.06280 
\bibitem[Markevitch \& Vikhlinin(2007)]{MV07} Markevitch, M., \& Vikhlinin, A.\ 2007, \physrep, 443, 1
\bibitem[Matsushita(2011)]{mat11} Matsushita, K.\ 2011, \aap, 527, A134 
\bibitem[McDonald et al.(2016)]{mcd16} McDonald, M., Bulbul, E., de Haan, T., et al.\ 2016, \apj, 826, 124 
\bibitem[Mernier et al.(2015)]{mer15} Mernier, F., de Plaa, J., Lovisari, L., et al.\ 2015, \aap, 575, A37 
\bibitem[Nagai et al.(2007)]{nag07} Nagai, D., Vikhlinin, A., \& Kravtsov, A.~V.\ 2007, \apj, 655, 98
\bibitem[Nelson et al.(2014)]{nel14} Nelson, K., Lau, E.~T., \& Nagai, D.\ 2014, \apj, 792, 25
\bibitem[Ohno et al.(2002)]{ohn02} Ohno, H., Takizawa, M., \& Shibata, S.\ 2002, \apj, 577, 658 
\bibitem[Ota et al.(2007)]{ota07} Ota, N., Fukazawa, Y., Fabian, A.~C., et al.\ 2007, \pasj, 59, 351 
\bibitem[Piffaretti \& Valdarnini(2008)]{pif08} Piffaretti, R., \& Valdarnini, R.\ 2008, \aap, 491, 71
\bibitem[Pinto et al.(2015)]{pin15} Pinto, C., Sanders, J.~S., Werner, N., et al.\ 2015, \aap, 575, A38
\bibitem[Randall et al.(2015)]{ran15} Randall, S.~W., Nulsen, P.~E.~J., Jones, C., et al.\ 2015, \apj, 805, 112 
\bibitem[Rasia et al.(2006)]{ras06} Rasia, E., Ettori, S., Moscardini, L., et al.\ 2006, \mnras, 369, 2013
\bibitem[Reiprich et al.(2009)]{rei09} Reiprich, T.~H., Hudson, D.~S., Zhang, Y.-Y., et al.\ 2009, \aap, 501, 899 
\bibitem[Ricker(2008)]{ric08} Ricker, P.~M.\ 2008, ApJS, 176, 293
\bibitem[Roediger et al.(2013)]{rod13} Roediger, E., Kraft, R.~P., Forman, W.~R., et al.\ 2013, \apj, 764, 60
\bibitem[Sanders et al.(2011)]{san11} Sanders, J.~S., Fabian, A.~C., \& Smith, R.~K.\ 2011, \mnras, 410, 1797
\bibitem[Sanders \& Fabian(2013)]{san13} Sanders, J.~S., \& Fabian, A.~C.\ 2013, \mnras, 429, 2727
\bibitem[Sarazin(1988)]{sar88} Sarazin, C.~L.\ 1988, X-Ray Emission from Clusters of Galaxies (Cambridge: Cambridge Univ. Press)
\bibitem[Schuecker et al.(2004)]{sch04} Schuecker, P., Finoguenov, A., Miniati, F., B{\"o}hringer, H., \& Briel, U.~G.\ 2004, \aap, 426, 387
\bibitem[Simionescu et al.(2012)]{sim12} Simionescu, A., Werner, N., Urban, O., et al.\ 2012, \apj, 757, 182
\bibitem[Smith et al.(2001)]{smi01} Smith, R.~K., Brickhouse, N.~S., Liedahl, D.~A., \& Raymond, J.~C.\ 2001, \apjl, 556, L91
\bibitem[Spitzer(1962)]{spi62} Spitzer, L.\ 1962, Physics of Fully Ionized Gases, New York: Interscience (2nd edition), 1962
\bibitem[Sugawara et al.(2009)]{sug09} Sugawara, C., Takizawa, M., \& Nakazawa, K.\ 2009, \pasj, 61, 1293 
\bibitem[Suto et al.(2013)]{sut13} Suto, D., Kawahara, H., Kitayama, T., et al.\ 2013, \apj, 767, 79 
\bibitem[Takahashi et al.(2014)]{tak14} Takahashi, T., Mitsuda, K., Kelley, R., et al.\ 2014, \procspie, 9144, 914425
\bibitem[Takizawa et al.(2010)]{tak10} Takizawa, M., Nagino, R., \& Matsushita, K.\ 2010, \pasj, 62, 951 
\bibitem[Tamura et al.(2011)]{tam11} Tamura, T., Hayashida, K., Ueda, S., \& Nagai, M.\ 2011, \pasj, 63, S1009 
\bibitem[Tamura et al.(2014)]{tam14} Tamura, T., Yamasaki, N.~Y., Iizuka, R., et al.\ 2014, \apj, 782, 38 
\bibitem[Turk et al.(2011)]{tur11} Turk, M.~J., Smith, B.~D., Oishi, J.~S., Skory, S., Skillman, S.~W., Abel, T., \& Norman, M.~L.\ 2011, \apjs, 192, 9
\bibitem[Walker et al.(2017)]{wal17} Walker, S.~A., Hlavacek-Larrondo, J., Gendron-Marsolais, M., et al.\ 2017, \mnras, 468, 2506 
\bibitem[Werner et al.(2009)]{wer09} Werner, N., Zhuravleva, I., Churazov, E., et al.\ 2009, \mnras, 398, 23
\bibitem[Wilms et al.(2000)]{wil00} Wilms, J., Allen, A., \& McCray, R.\ 2000, \apj, 542, 914
\bibitem[Zhuravleva et al.(2013)]{zhu13} Zhuravleva, I., Churazov, E., Sunyaev, R., et al.\ 2013, \mnras, 435, 3111
\bibitem[Zhuravleva et al.(2014)]{zhu14} Zhuravleva, I., Churazov, E., Schekochihin, A.~A., et al.\ 2014, \nat, 515, 85
\bibitem[Zhuravleva et al.(2015)]{zhu15} Zhuravleva, I., Churazov, E., Ar{\'e}valo, P., et al.\ 2015, \mnras, 450, 4184
\bibitem[Zhuravleva et al.(2016)]{zhu16} Zhuravleva, I., Churazov, E., Ar{\'e}valo, P., et al.\ 2016, \mnras, 458, 2902 
\bibitem[ZuHone et al.(2010)]{zuh10} ZuHone, J.~A., Markevitch, M., \& Johnson, R.~E.\ 2010, \apj, 717, 908
\bibitem[ZuHone et al.(2013)]{zuh13} ZuHone, J.~A., Markevitch, M., Brunetti, G., \& Giacintucci, S.\ 2013, \apj, 762, 78
\bibitem[ZuHone et al.(2014)]{zuh14} ZuHone, J.~A., Biffi, V., Hallman, E.~J., et al.\ 2014, arXiv:1407.1783
\bibitem[ZuHone et al.(2015)]{zuh15} ZuHone, J.~A., Kunz, M.~W., Markevitch, M., Stone, J.~M., \& Biffi, V.\ 2015, \apj, 798, 90
\bibitem[ZuHone et al.(2016)]{zuh16} ZuHone, J.~A., Miller, E.~D., Simionescu, A., \& Bautz, M.~W.\ 2016, \apj, 821, 6 (Z16)

\end{thebibliography}
\end{document}